\documentstyle[12pt]{article}
\def\ct{canonical transformation}

\def\mss{minisuperspace}
\newcommand{\half}{\frac{1}{2}}

\def\fraction#1#2{{\textstyle{#1\over#2}}} \def\fr{\fraction}
 
\def\d{\partial}

\def\journalfont{\it}         
\def\jou#1{{\journalfont #1}}
\def\am{\jou{   Ann.\ Math.)}}
\def\cmp{\jou{  Commun.\ Math.\ Phys.}}
\def\cqg{\jou{  Class.\ Quantum Grav.}}
\def\grg{\jou{  Gen.\ Relativ.\ Grav.}}
\def\jmp{\jou{  J.\ Math.\ Phys.}}
\def\pr{\jou{   Phys.\ Rev.}}

\def\Lie{{\cal L}}

\def\cone{\hbox{$\Lie^+$}}
\def\halfcone{\hbox{$\Lie^+_R$}}
      \def\ket#1{|\, #1\, \rangle}
      
\def\IP#1#2{\langle\, #1\, |\, #2\, \rangle}  
\def\Div#1{{\rm Div}_{#1}}                

\def\ps{phase space}
\def\rd{{\rm d}}

\def\ch#1{(\cosh 2\sqrt{3}\tbp)^{#1}}
\def\sh{\sinh(2\sqrt{3}\tbp)}

\newcommand{\bz}{\beta^0}
\newcommand{\bp}{\beta^+}             \newcommand{\bm}{\beta^-}
\newcommand{\pz}{p_0}
\newcommand{\pp}{p_+}                 \newcommand{\pmm}{p_-}

\newcommand{\bbz}{\bar\beta^0}
\newcommand{\bbp}{\bar\beta^+}        \newcommand{\bbm}{\bar\beta^-}
\newcommand{\bpz}{{\bar p}_0}
\newcommand{\bpp}{\bar{p}_+}          \newcommand{\bpm}{\bar{p}_-}

\newcommand{\tb}{\tilde\beta}         \newcommand{\tbz}{\tilde\beta^0}
\newcommand{\tbp}{\tilde\beta^+}      \newcommand{\tbm}{\tilde\beta^-}
\newcommand{\tp}{\tilde{p}}           \newcommand{\tpz}{{\tilde p}_0}
\newcommand{\tpp}{{\tilde p}_+}       \newcommand{\tpm}{{\tilde p}_-}

\newcommand{\be}{\begin{equation}}
\newcommand{\ee}{\end{equation}}
\newcommand{\bea}{\begin{eqnarray}}
\newcommand{\eea}{\end{eqnarray}}

\catcode`@=11

\newdimen\jot \jot=3pt
\newskip\z@skip \z@skip=0pt plus0pt minus0pt
\newdimen\z@ \z@=0pt 
\dimendef\dimen@=0

\def\m@th{\mathsurround=\z@}

\def\ialign{\everycr{}\tabskip\z@skip\halign} 

\def\openup{\afterassignment\@penup\dimen@=}
\def\@penup{\advance\lineskip\dimen@
  \advance\baselineskip\dimen@
  \advance\lineskiplimit\dimen@}

\def\eqalign#1{{
\null\,\vcenter{\openup\jot\m@th
  \ialign{\strut\hfil$\displaystyle{##}$&$\displaystyle{{}##}$\hfil
      \crcr#1\crcr}}\,} }
\catcode`@=12   

\begin{document}
\bibliographystyle{kr}  


\begin{center}
{\Large {\bf Minisuperspaces: Observables and Quantization}}
\end{center}
\medskip
{\centerline{SU-GP-92/2-6, gr-qc/9302027}}
\bigskip

\begin{center}
{\large Abhay Ashtekar$^{\dagger}$, Ranjeet Tate$^{\dagger\star}$
  and Claes Uggla$^{\dagger \ddagger}$}
\end{center}
\begin{flushleft}
{\small $\dagger$ Physics Department, Syracuse University,
Syracuse, NY 13244-1130, USA} \\
{\small $\star$ Physics Department, University of California, Santa Barbara,
CA 93106-9530, USA} \\
{\small $\ddagger$ Department of Physics, Lule\aa\  University of Technology,
S-951 87 Lule\aa, Sweden}
\end{flushleft}

\smallskip

{\small
\begin{center}  {\large {\bf Abstract}} \end{center}
A canonical transformation is performed on the phase space of a number
of homogeneous cosmologies to simplify the form of the scalar
(or, Hamiltonian) constraint. Using the new canonical coordinates, it is
then easy to obtain explicit expressions of Dirac observables, i.e.\ phase
space functions which commute weakly with the constraint. This, in turn,
enables us to carry out a general quantization program to completion. We
are also able to address the issue of time through ``deparametrization''
and discuss physical questions such as the fate of initial singularities
in the quantum theory. We find that they persist in the quantum theory
{\it inspite of the fact that the evolution is implemented by a 1-parameter
family of unitary transformations}. Finally, certain of these models admit
conditional
symmetries which are explicit already prior to the canonical transformation.
These can be used to pass to quantum theory following an independent avenue.
The two quantum theories --based, respectively, on Dirac observables in the
new canonical variables and conditional symmetries in the original ADM
variables-- are compared and shown to be equivalent.}





\section{Introduction}

Quantum general relativity has a number of peculiar features which are
not encountered in quantum theories of non-gravitational interactions:
presence of ``dynamical'' constraints; diffeomorphism invariance and
the consequent absence of a background geometry; nonlinearities which
make the structure of the effective configuration space topologically
complicated; the absence of suitable symmetries to single
out the vacuum and select the Hermitian scalar product; and, the
difficulty of interpreting the resulting mathematical framework in
simple physical terms. Over the past six years, a nonperturbative
approach has been developed to address these issues systematically
(see, e.g.\ \cite{newbook,cr:rev,aa:rev}). In particular, there now exists
\cite{newbook,aa:rst} a general quantization program which is adapted to
the peculiarities of general relativity. Unfortunately, however, for the
full, untruncated theory, several steps of the program are yet to be
completed.

It is therefore desirable to apply the program to simpler, truncated
models (see e.g.\ \cite{aa:rst,rst:thesis}) both to test its viability
and to gain insight into the type of techniques that will be needed in
the full theory. One such model is provided by 3-dimensional general
relativity \cite{2+1}. This model has taught us several interesting
lessons, both conceptual and technical (see chapter 17 in
\cite{newbook}).  The purpose of this paper is to continue
investigations in the same spirit by examining a class of ``solvable''
spatially homogeneous cosmologies.
These are homogeneous cosmologies which admit additional symmetries.
In the classical theory, the presence of these symmetries enables one
to integrate the field equations completely. We will see that their
presence also simplifies the task of quantization: We will be able to
carry out the general program of \cite{newbook,aa:rst} to completion.
Furthermore, these models will enable us to explore certain aspects of
the program which could not be analysed in 3-dimensional general
relativity.

The key simplification that allows one to complete the quantization
program is the following: in these models one can perform a canonical
transformation on the classical phase space to drastically simplify
the expression of the scalar (also known as the Hamiltonian)
constraint of geometrodynamics. In the new canonical variables, the
potential term in the scalar constraint {\it disappears entirely!}.
Furthermore, the supermetric in the kinetic term is {\it flat}. The
only remnant of the potential term of the usual ADM
\cite{adm} variables is in the ranges of permissible values of the new
canonical coordinates, i.e., in the global topology of the constraint
surface. Now, in full general relativity, one can again significantly
simplify the form of the constraints by performing (quite different)
canonical transformations which too, in particular, remove the
potential term from the scalar constraint (see e.g.
\cite{aa:nv,rst:poly}).  Although the origins and the forms of these
canonical transformations are quite different from the one used in
this paper, there is nonetheless some qualitative similarity between the
situations. One can exploit it to gain some insight into the full theory.

In particular, it is of considerable interest to understand the effect
of such canonical transformations on quantization. Are the quantum
theories based on the old and the new canonical coordinates --or, more
precisely, on polarizations of the phase space naturally adapted to
the old and the new canonical coordinates-- equivalent? In full
general relativity, one cannot answer the question at the present
stage since the quantization program remains incomplete in both old
and new variables: in the older ADM variables no solution to the
quantum constraint is known while in the newer connection
variables \cite{aa:nv,newbook}, although a family of solutions {\it is}
known, the Hilbert space structure on the space of these solutions is yet
to be determined. For some of the models under consideration, on the
other hand, both programs can be completed so that a comparison {\it is}
possible.

The overall situation can be summarized as follows. First, by exploiting
the simplicity of the form of the scalar constraint in the new variables,
one can obtain the general solution to its quantum version.
Furthermore, one can find a complete set of Dirac
observables ---functions on phase space whose Poisson brackets with
the constraint vanish weakly. Now, in the quantization program of
\cite{newbook,aa:rst}, one selects the inner product on physical
states --i.e.\ on solutions to quantum constraints-- by demanding that
the real Dirac observables be promoted to self-adjoint operators in
the quantum theory. In all the models under consideration, this strategy
works and provides us with a distinguished $\star$-representation of the
algebra of Dirac observables. Second, in the traditional ADM variables,
some of these models also admit what are known as conditional symmetries
\cite{kk:jmp}. These are 1-parameter families of diffeomorphisms on the
effective configuration space whose action commutes with the quantum
scalar constraint, i.e., the Wheeler-DeWitt operator. Thus, if we pass
to the quantum theory using the traditional quantization method, we
acquire 1-parameter symmetry groups acting on the space of physical
states. By requiring that this action be unitary, one can uniquely
``Hilbertize'' the space of physical states thereby achieving the goal
of quantum geometrodynamics \cite{atu:I}.  This second procedure
relies on the existence of symmetries on the effective configuration
space of the model; unlike in the strategy followed
using new variables, there is no need to isolate a {\it complete} set
of Dirac observables. It is therefore not clear a priori that the two
quantum theories --based on the new and the old canonical variables--
would be equivalent. We will show that they in fact are, although some
of the issues that arise are subtle and require care. This analysis
illustrates the type of issues that are likely to arise when we explore
the effect of the canonical transformation \cite{aa:nv} on quantization of
the full theory.

The completion of the various steps in the program provides us with a
Hilbert space of physical states and a complete set of physical
observables.  Yet, the theory remains difficult to interpret
physically because one is left with what is often called a ``frozen
formalism'' in which nothing happens.  This comes about because in
the classical theory, dynamics is generated by a constraint. To
illustrate this point, let us consider a free (relativistic) particle
of mass $m$ in Minkowski space. In the classical Hamiltonian
description, configuration space is the 4-dimensional Minkowski space,
phase space is the cotangent bundle over it, and dynamics is governed
by the constraint $P\cdot P +m^2 =0$. To quantize the system it is
simplest to work in the momentum representation. The physical states
are then square-integrable functions on the (future) mass shell. These
are all annihilated by the quantum constraint. There is no notion of
time or of evolution; nothing happens. The picture we obtain by
carrying out the quantization program for the spatially homogeneous
models under consideration is completely analogous. Recall however,
that in the case of the relativistic particle, one can cast the
quantum theory in another form in which dynamics does appear. One can
simply consider the position representation and rewrite the quantum
constraint as the (positive frequency component of the) Klein Gordon
equation. The wave function is then seen to evolve in time: the
quantum constraint equation simply reduces to the evolution equation
for quantum states. It turns out that the simplicity of the
homogeneous models under consideration enables us to treat the issue
of time in a completely analogous fashion. More precisely, we will be
able to deparametrize \cite{kk:qg2,newbook} the theory explicitly and
show that in the quantum theory, the scalar constraint reduces to a
Schr\"odinger evolution equation. This in turn will enable us to
analyse how various physically interesting observables evolve in time.
In particular, we will be able to explore the fate of singularities in
the quantum theory.  More generally, this framework will enable us to
interpret the mathematical framework of the quantum theory in direct
physical terms.

The plan of the paper is as follows. Section 2 is devoted to
preliminaries.  We recall general facts about homogeneous cosmologies
and single out the models to be discussed in detail. In section 3, we
present the canonical transformation which removes the potential term
from the scalar constraint. In section 4 we carry out the quantization
program using the new canonical variables. We discuss the physical
interpretation of the resulting mathematical structure in section 5.
In section 6, we pursue quantization in the old canonical variables
(with the potential term in the expression of the scalar constraint)
via the alternate strategy of using conditional symmetries. While the
resulting quantum theory is difficult to interpret because of the lack
of (explicit expressions of) Dirac observables, it is mathematically
complete from the traditional viewpoint. We conclude section 6 by
comparing this mathematical framework with that developed in section
4. In section 7 we summarize both the overall picture and the lessons
that can be drawn from this analysis.

The paper thus contains several related but distinct results. Readers
familiar with the basic facts of spatially homogeneous models can
proceed directly to the summary at the end of section 2. Readers whose
primary interest lies in the technical and conceptual problems of
quantization rather than in Bianchi models can skip section 2 and most
of section 3 and proceed directly to the summary of the Hamiltonian
structure presented in the last paragraph of section 3. Finally, readers
whose primary interest lies in the in the issue of time and dynamics in
(canonical) quantum gravity and who are familiar with the Hamiltonian
structure of the Bianchi I model can proceed directly to section 5.

\section{Mathematical Preliminaries}

In this paper, we will consider diagonal, spatially homogeneous models
which admit intrinsic, multiply transitive symmetry groups. For
completeness, in this section we will specify the meaning of various
terms appearing in the definition of this class and place this class
in the general context of spatially homogeneous space-times. However,
this material is not needed directly in the main part of the paper.

A spacetime is said to be {\it spatially homogeneous} if it admits a
foliation by space-like sub-manifolds such that the isometry group of
the 4-metric acts on each leaf transitively. If the action of the
isometry group is multiply transitive and if there is no subgroup
whose action is simply transitive, the spacetime is of {\it
Kantowski-Sachs} type. If on the other hand, the isometry group admits
a (not necessarily proper) subgroup which acts simply transitively on
each leaf, the spacetime is said to be of {\it Bianchi} type. In this
case, one focuses on the subgroup --which is necessarily
3-dimensional-- and further classifies space-times using the properties
of the corresponding Lie algebras. If the trace
$C^a{}_{ba}$ of structure constants $C^a{}_{bc}$ of the Lie algebra
vanishes, the space-time belongs to {\it Bianchi Class A} while if it
does not vanish, it belongs to {\it Bianchi class B}
\cite{m:ani}.

A spatially homogeneous 4-metric is said to be {\it diagonal} if it
can be written in the form:
\be
   ds^2 = -(N(t))^2dt^2 + \sum_{a=1}^3 g_{aa}(t) (\omega^a)^2\ ,
\ee
where $N(t)$ is the lapse function and $\omega^a$ is a basis of spatial
1-forms which are left invariant by the action of the isometry group.
One can always change the time-coordinate $t$ to proper-time so that
the coefficient of the first term is simply $-1$. The diagonal models
are then characterized by the three components $g_{aa}$ which are
functions only of time. A key issue, however, is whether the diagonal
form of the metric is compatible with the vacuum field equations. This
is the case for models for which the vector (or, the diffeomorphism)
constraint is identically satisfied and only the scalar (or, the
Hamiltonian) constraint remains to be imposed. In this paper, we will
restrict ourselves to this class of models since they admit a
Hamiltonian formulation, which is the point of departure for canonical
quantization. In the case of Kantowski-Sachs metrics and the class A
models, the vector constraint is identically satisfied; they belong to the
class under consideration. For class B models, on
the other hand, compatibility with field equations is not automatic:
It is only restricted versions of type III, V and VI models (in which
only two of the three metric coefficients are independent) that are
both diagonal and satisfy the vector constraint identically. To
summarize: the class of diagonal models compatible with the vacuum
field equations consists of class A Bianchi models, in which the
minisuperspaces will be 3-dimensional; Kantowski-Sachs models in which
they will be 2-dimensional (since two of the $g_{aa}$ are always equal
in these models) and certain class B Bianchi models in
which they will be again 2-dimensional. Thus these models have either
2 or 1 true degree of freedom.

Misner \cite{m:mini} has introduced a very useful parametrization of
the diagonal spatial metric:
\be \label{eq:param}
g_{aa} = e^{2\beta^a}\ ,\qquad
 \left( \begin{array}{c}
        \beta^1 \cr
        \beta^2 \cr
        \beta^3 \cr
        \end{array} \right) =
 \left( \begin{array}{ccc}
         1 & 1 & \sqrt3 \cr
         1 & 1 & -\sqrt3 \cr
         1 & -2 & 0 \cr
        \end{array} \right)
 \left( \begin{array}{c}
         \beta^0 \cr
         \beta^+ \cr
         \beta^- \cr
        \end{array} \right)\ .
\ee
We will see that the further restrictions we have to impose to arrive
at the class of models which are of interest in this paper can be
expressed concisely in terms of the parameters $(\beta^0 ,\beta^+,
\beta^- )$.

Before discussing these restrictions, however, let us explore class A
models in a little more detail. Since the trace of the structure
constants vanishes for these models, they can be expressed entirely in
terms of a symmetric, second rank matrix $n^{ab}$ \cite{j:uni}:
\be
 C^a{}_{bc} = \epsilon_{mbc}n^{ma},
\ee
where $\epsilon_{mbc}$ is the completely anti-symmetric symbol. The
signature of $n^{am}$ can then be used to divide the class A models
into various types: If $n^{ab}$ vanishes identically, we have Bianchi
type I; if it has signature $(0,0,+)$, we have type II; signature
$(+,-,0)$ corresponds to type VI$_0$; $(+,+,0)$ corresponds to
VII$_0$; $(+,+,-)$ to type VIII; and, $(+,+,+)$ to type IX.

For Bianchi class B models, the trace $C^a{}_{ba} =: a_b$ does not
vanish and implies a decomposition of the structure constants of the
form:
\be
 C^a{}_{bc} = \epsilon_{mbc}n^{ma} + a_{[b}\delta^a{}_{c]},
\ee
where $n^{ab}$ is again symmetric but now satisfies the constraint
$n^{ab}a_b = 0$. The models are now classified by the signature of
$n^{ab}$ and --if the zero-eigenvector $a_b$ of $n^{ab}$ is
non-degenerate-- in addition by the value of the constant $h$ defined
via:
\be
 a_ma_n = \textstyle{h\over 2} \epsilon_{mab}\epsilon_{ncd}n^{ac}n^{bd}.
\ee
For type V metrics $n^{ab}$ vanishes. For type III metrics, $n^{ab}$
has signature $(+, -, 0)$ and $h$ equals $-1$. We shall not specify
values of these parameters for the remaining class B models because we
will not need them.

We can now continue the specification of conditions to arrive at the
class of space-times of interest in this paper.  The next restriction
is to {\it multiply transitive} diagonal models.  The Kantowski-Sachs
models obviously belong to this class. Among Bianchi types, the
condition of multiple transitivity leads to a further restriction
since a generic Bianchi model is only simply transitive. In these
models, if an additional Killing vector exists, it is always a
rotation \cite{j:uni,kra:ex} and these space-times are referred to in
the literature as {\it locally rotationally symmetric (LRS) models}
\cite{e:lrs}. In each of the Bianchi types I, II, VII$_0$, VIII and IX,
one can obtain a LRS model simply by setting the Misner parameter
$\beta^-$ equal to zero. (Thus, among class A models, only type VI$_0$
fails to admit a multiply transitive sub-family.) It turns out,
however, that the family of LRS type VII$_0$ models coincides with the
family of LRS type I models.  Therefore, we need only consider LRS
families associated with types I, II, VIII and IX. Finally there is a
family of models admitting a mutliply transitive symmetry group among
the class B space-times. This family consists of the LRS type III
models and the isotropic type V models. The LRS type III models are
obtained by setting $\beta^-$ equal to zero while the isotropic type V
models are obtained by setting both $\beta^\pm$ to zero.

In this paper, however, we consider models which are somewhat {\it
more general}\ than the diagonal multiply transitive ones: we need the
multiple transitivity to hold {\it only intrinsically}. In the
Kantowski-Sachs case, this loosening of the restriction makes no
difference. In the Bianchi models, however, it does: we require the
additional symmetry to be a Killing field only of the 3-metric
intrinsic to the homogeneous slices and not necessarily of the full
4-metric.  Clearly, all multiply transitive models are also
intrinsically multiply transitive. However, we now acquire additional
models: diagonal Bianchi types I and II without any further
restrictions and diagonal type V models with the restriction that
$\beta^+$ is set to zero in order to make the vector constraint vanish
identically. Types I and II belong to class A while V belongs to class
B. The \mss s are 3-dimensional in types I and II and
2-dimensional in type V \cite{ruj:vac}.

To summarize then, the diagonal, intrinsically multiply transitive
(DIMT) models are the following: Bianchi types I and II; sub-families
of Bianchi types III, VIII and IX defined by $\beta^- =0$;
Kantowski-Sachs space-times; and Bianchi type V models with $\beta^+ =
0$. For Bianchi types I and II, the \mss s are 3-dimensional,
parametrized by $ \beta^0, \beta^+ $ and $\beta^-$ in the Misner
scheme; for Bianchi type V, the 2-dimensional \mss\ is parametrized by
$\beta^0$ and $\beta^-$; and, for all the remaining models --which are
diagonal, multiply transitive (DMT)-- the \mss s are 2-dimensional,
parametrized by $\beta^0$ and $\beta^+$. In the remainder of this
paper, we restrict ourselves to these models. Thus, when we speak of
type VIII or type IX, for example, unless otherwise specified, we will
mean LRS type VIII and LRS type IX.

By exploiting the additional symmetries, one can explicitly solve the
vacuum equations in all DIMT models. For details and especially for
explicit expressions for the invariant 1-forms $\omega^a$ (defined in
(1)) for all of these models, see \cite{kra:ex,j:uni,ruj:vac}.

\section{Canonical Transformations}

This section is divided into three parts. In the first, we introduce the
general framework. In the Misner variables, the scalar constraint has the
familiar form of a sum of a kinetic term and a potential term. The idea
is to perform a canonical transformation so that --when expressed in terms
of the new canonical variables-- the potential term disappears entirely.
This is achieved in the second and the third sub-sections. The resulting
Hamiltonian description is remarkably simple and serves as the point of
departure for quantization in the next section.

\subsection{Hamiltonian framework} \label{sec:ham}

One can use the ADM procedure \cite{adm,j:uni} to arrive at the Hamiltonian
formulation of all DIMT models. The configuration spaces can be labelled by
the Misner parameters ---collectively denoted as $\beta^A$ in what follows---
which take values in $(-\infty, \infty)$; the spaces are topologically trivial.
(The index $A$ will take values $0,+,-$ in type I and II models, $0,-$ in type
V and $0,+$ in the remaining (i.e., DMT) models.) Following the terminology
commonly used, we will refer to them as minisuperspaces. The phase spaces
are cotangent bundles over these configuration spaces. We will denote the
momenta conjugate to $\beta^A$ by $p_A$. Thus, the fundamental Poisson
bracket relations are $\{\beta^A, p_B\} = \delta^A_B$.

As noted above, in all these models, the vector constraints are
automatically satisfied and one is left only with the scalar
constraint. To simplify calculations, one usually chooses the ``Taub
time gauge'', i.e., one chooses the lapse function $N_T=
12\exp{3\beta^0}$ \cite{taub51}. (This time gauge is also known as
Misner's supertime gauge \cite{m:mini}.) To discuss the structure of
the resulting constraint function, it is convenient to treat the type
V models separately. With the Taub gauge choice the scalar constraint
for the type V models takes the form \cite{ruj:vac}:
\be\eqalign{  \label{eq:scalv}
  C_T &\equiv C_0 + C_-=0\ ,\cr
  C_0 &= -\fr12 p_0{}^2 + k_0e^{4\bz}\ , \qquad C_- = \fr12 p_-{}^2\
,\cr}
\ee
where $k_0 = 72$.  For the remaining models we have \cite{j:uni,ujr:hid}:
\be\eqalign{  \label{eq:scal}
  C_T &= \fr12 \eta^{AB} p_A p_B + U_T = 0\ ,\cr
  U_T &= k_0 e^{2(2\bz - \bp)} + k_+ e^{4(\bz - 2\bp)}\ ,\cr}
\ee
where $\eta^{AB}$ is $diag(-1,1,1)$ for type I and II models and
$diag(-1,1)$ for the remaining (i.e.\ DMT) models; and the values of
the constants $k_0$, $k_+$ and $k_-$ (defined below) characterizing
the different models are given in the following table:
\vskip0.5truecm
\renewcommand{\arraystretch}{2}
\centerline{
 \begin{tabular}{||c||c|c|c|c|c|c||} \hline
        & I  & II & VIII &   IX  &   KS  &  III  \\\hline
  $k_0$ & 0  & 0  &  24  & $-24$ & $-24$ &   24     \\\hline
  $k_+$ & 0  & 6  &  6   &    6  &    0  &    0     \\\hline
  $k_-$ & 1  & 1  &  0   &    0  &    0  &    0     \\\hline
 \end{tabular}
}
\vskip0.5truecm

Finally, for the non-type V models, it is convenient to make a linear point
transformation to cast the scalar constraint in a form that will be
particularly useful. Set \cite{j:uni}
\be \label{eq:bb}
   (\bbz, \bbp , \bbm) = \fr{1}{\sqrt{3}}(2\bz - \bp, -\bz + 2\bp, \sqrt3 \bm)
   \ .
\ee
Then, the scalar constraint can be re-expressed as:
\be\eqalign{ \label{eq:solvc}
  C_T &\equiv C_0 + C_+ + C_- = 0\ , {\rm where}\cr
  C_0 &= -\fr12 \bpz{}^2 + k_0 e^{2\sqrt3 \bbz}\ ,\qquad
  C_+ =  \fr12 \bpp{}^2 + k_+ e^{-4\sqrt3 \bbp}\ ,\qquad
  C_- =  \fr12 k_- \bpm{}^2\ .\cr
}\ee

Thus, the scalar constraint for all DIMT models is separable, a fact
which underlies the ``solvability'' of these models. This property
will enable us in the next two sub-sections to perform the canonical
transformation which will drastically simplify the form of the
constraint.

\subsection{Cases with non-positive $k_0$}

When $k_0$ is non-positive, the constraint is a sum of terms of the form:
\be \label{eq:sep}
   C^{(1)} = \epsilon\fr12 ({\bar p}_A{}^2 + a_A{}^2e^{2d_A{\bar
        \beta}^A})\ ,
\ee
with no sum over $A$. The constant coefficients $a_A, d_A$ are given
by: $a_-=0$; $a_A^2=2|k_A|$ for $A=0,+$; $d_0=\sqrt3$ and
$d_+=-2\sqrt3$. Finally, $\epsilon$ takes the values $\pm 1$. We now
want to define new coordinates $\tb^A$ and momenta $\tp_A$ so that
terms $C^{(1)}$ in (\ref{eq:sep}) take the form $C^{(1)} =
\epsilon\half\tp_A{}^2\, $ so that the potential term disappears
completely.  Clearly, if $a_A =0$, the corresponding term in the
expression of the scalar constraint is already of the desired form
whence we can and will simply set $\tb^A= \bar\beta^A$ and $\tp_A
=\bar{p}_A$; no canonical transformation is needed in {\it that}
$(\bar{\beta}^A,\bar{p}_A)$ plane. Therefore, in this and the next
sub-section, we will consider in detail only those cases in which $a_A
\neq 0$, and assume, without loss of generality, that $a_A$ is
positive.

The required canonical transformation is easy to obtain. Let us begin by
setting
 \be  \label{eq:ct1}
  \tp_A = \sqrt{{\bar p}_A{}^2 + a_A{}^2e^{2d_A{\bar\beta}^A}}.
 \ee
Note that by definition, $\tp_A > 0$.
To find the corresponding canonically conjugate $\tb^A$,
we have to solve an elementary differential equation. The result is:
 \be   \label{eq:ct1p}
   \tb^A = \frac1{d_A}\left(\ln[-{\bar p}_A + \sqrt{
           {\bar p}_A{}^2 + a_A{}^2e^{2d_A{\bar\beta}^A}}] -
           \ln[a_Ae^{d_A{\bar\beta}^A}]\right) \ .
 \ee
Thus (\ref{eq:ct1}) and (\ref{eq:ct1p}) together define a \ct\ which yields
a $C^{(1)}$ of the desired form. The inverse of this \ct\ is given by
 \be
   a_Ae^{d_A{\bar\beta}^A} = \tp_A/\cosh(d_A\tb^A)\ ,\qquad
   {\bar p}_A = -\tp_A \tanh(d_A\tb^A) \ ,
 \ee
from which we can see that the canonical transformation is globally defined
on the \ps, since $(\bar\beta^A, {\bar p}_A)$ take all values.

Note that $C_+$ is always of the form $C^{(1)}$, with $\epsilon=1$.
Thus the above \ct\ (\ref{eq:ct1}, \ref{eq:ct1p}) leads to
the desired form for this term, $C_+ = \fr12 \tpp{}^2$, {\it and}\ a
non-holonomic constraint ${\tilde p}_+ > 0$. This last constraint is
important since it is now the only remnant of the potential term in the
barred variables.

As far as $C_0$ is concerned, in type I and II models, we can simply set
$\bbz =\tbz$ and $\bpz=\tpz$, since $k_0$ --and hence $a_0$-- vanishes
in these cases. As the description stands, there is no restriction on the
sign of $\bpz$ and hence of $\tpz$. However, if we flip the sign, the
dynamical trajectories in the phase space --and hence space-time geometries
they define-- remain unaltered. What changes is the sign of the affine
parameter along the trajectories in the phase space, or equivalently,
the convention regarding future versus past evolution in the physical
space-time picture. Keeping both signs is therefore redundant.%
\footnote{Even if one ignores this redundancy and keeps both signs of $\tpz$
 in the classical theory, the requirement that one restrict oneself to an
 {\it irreducible} representation of the algebra of  Dirac observables
 forces one to choose one {\it or} the other sign in quantum theory.}
In order to make the Hamiltonian description parallel to the textbook
discussion of the relativistic particle, we will choose $\tpz \ge 0$.%

When $k_0<0$, i.e.\ for the KS and the LRS type IX models, $C_0$ is again of
the form $C^{(1)}$, with $\epsilon=-1$. For these cases, it follows from
(\ref{eq:ct1}) that we
again have $\tpz > 0$. This leads to the form $C_0 = -\fr12 {\tilde p}_0{}^2$
{\it and} the restriction ${\tilde p}_0 > 0$. Thus we have shown that
one can use the canonical transformation (\ref{eq:ct1}, \ref{eq:ct1p}) to
transform the constraint (for all DIMT models except type V, and LRS types III
and VIII, which we will discuss below) to the form
\be  \label{eq:rfp}
   C_T = \fr12 \eta^{AB}\tp_A \tp_B = 0\ .
\ee
Thus, as was desired, now the constraint contains only a kinetic piece
quadratic in momenta; the potential term has been eliminated. Furthermore,
the metric defined by the kinetic term is just the flat Minkowski metric!
Thus, locally in phase space, the dynamics of any of these models is the
same as that of any other and is furthermore indistinguishable from that
of a massless, relativistic, free particle in a two or three dimensional
Minkowski space. The canonical transformation has essentially enabled us
to pass to the ``action angle type'' variables appropriate to each model.

However, in each of these models, the information in the potential terms
is now essentially coded in the {\it global} structure of the phase space.
Due to the presence of the non-holonomic constraints on the $\tp_A$, one
can no longer consider them as momenta. However, in the $(\tb^A,\tp_A)$
coordinates, the \ps\ still has the structure of a cotangent bundle over
the space coordinatized by $\{\tp_A\}$. It is therefore convenient to
regard the $\tp_A$ as the configuration variables and the $\tb^A$ as the
corresponding momenta. The holonomic as well as the non-holonomic constraints
restrict only the configuration variables $\tp_A$. The restricted manifolds
are given by:
\begin{itemize}
  \item Bianchi type I: Future ($+$) light cone in 3 dimensions ($^3\!\cone$).
  \item LRS type I: Future light cone in 2 dimensions ($^2\!\cone$),
        obtained by setting $\beta^- = {\tilde \beta}^- =0$ in $^3\!\cone$.
  \item Bianchi type II: Right $(R)$ half of the future light cone in 3
        dimensions
        ($^3\!\halfcone$); since only the right half of $^3\!\cone$ is allowed
        due to the additional non-holonomic constraint $\tp_+ > 0$.
  \item LRS type II: Right half of the future light cone
        in 2 dimensions ($^2\!\halfcone$), obtained by setting
        ${\tilde \beta}^- = 0$ in $^3\!\halfcone$. (Note that this is just
        half the real line.)
  \item KS: Future light cone in 2 dimensions, $^2\!\cone$, as for LRS type I.%
\footnote{Strictly speaking, the origin should be excluded for the KS models.
 However, since the presence or absence of the solitary boundary  point makes
 only a trivial difference in the quantum theories under consideration, we will
 no longer draw a distinction.}
  \item LRS type IX: Right half of the future light cone in 2 dimensions
        $^2\!\halfcone$, as for LRS type II.
\end{itemize}
\goodbreak

\subsection{Cases with positive $k_0$}

The constant $k_0$ is positive in the following models: type V, LRS types III
and VIII. In these cases, the part $C_0$ of the scalar constraint is of the
form:
\be
   C^{(2)} = -\fr12 ({\bar p}_0{}^2 - a_0{}^2e^{2d_0{\bar \beta}^0})\ ,
\ee
where $a_0^2 = 2k_0$ in all cases,  $d_0 = 2$ in type V, and, $d_0=\sqrt{3}$
in the remaining cases. Therefore, the canonical transformations which
cast $C_0$ in the form $C_0 = -\frac12 \tp_0^2$ are now:
 \be \label{eq:ct2} \eqalign{
   \tp_0 &= \sqrt{{\bar p}_0{}^2 - a_0{}^2e^{2d_0{\bar\beta}^0}}\ ,\cr
   \tb^0 &= \frac1{d_0}\left(\ln[{\bar p}_0 - \sqrt{
           {\bar p}_0{}^2 - a_0{}^2e^{2d_0{\bar\beta}^0}}] -
           \ln[a_0e^{d_0{\bar\beta}^0}]\right) \ , \cr
 }\ee
and the inverse transformation assumes the form:
\be
  a_0e^{d_0{\bar\beta}^0} = \tp_0/\sinh(d_0\tb^0)\ ,\qquad
  {\bar p}_0 = \tp_0 \coth(d_0\tb^0) \ .
\ee
Note that there are regions of the \ps\ in which ${\bar p}_0{}^2 -
a_0{}^2e^{2d_0{\bar\beta}^0}$ is negative.  In these regions the new
``coordinate'' $\tp_0$, e.g., is imaginary. However, there exists a
neighborhood of the constraint surface in which ${\bar p}_0{}^2 -
a_0{}^2e^{2d_0{\bar\beta}^0}$ is everywhere positive, and thus $\tp_0$
is real. Furthermore, because of the simple form of the constraint, without
loss of generality, in the quantization procedure we will be able to
restrict ourselves to this neighborhood. Finally, on the
constraint surface, there is an additional non-holonomic constraint
$\tp_0 \geq 0$.

In all these cases --types V, LRS III and LRS VIII models-- $C_{+}$
and $C_{-}$, when not already in the desired form
$+\half\bar{p}_\pm^2$, continue to be of the form $C^{(1)}$ whence the
canonical transformations in the $(\tbp,\tpp)$ and $(\tbm, \tpm)$
planes are the same as those given in the previous subsection
(equations \ref{eq:ct1}, \ref{eq:ct1p}).  As before, the tilde
coordinates continue to exhibit the phase space as a cotangent bundle
and the natural configuration space is coordinatized by the
$\{\tp_A\}$.  The constraints again restrict only the configuration
space. We are led to the following list:
\begin{itemize}
   \item type V: Future light cone in 2 dimensions,
         $^2\!\cone$, as for LRS type I.
   \item LRS type III: Future light cone in 2 dimensions,
         $^2\!\cone$, as for LRS type I.
   \item LRS type VIII: Right half of the future light cone in 2 dimensions,
         $^2\!\halfcone$, as for LRS type II.
\end{itemize}

To summarize%
\footnote{We have restricted ourselves to vacuum space-times in this
 paper. However, one can sometimes can add sources to the DIMT models also
 which lead to separable constraints that can be transformed to the form
 (\ref{eq:rfp}). For example, one can add a cosmological constant $\Lambda$
 to the Bianchi type I models. This leads to a potential term $24\Lambda
 e^{6\bz}$ in the Taub time gauge; a term which is easily absorbed by a
 canonical transformation of the above type.},
the phase space dynamics for {\it all} the DIMT models has been reduced to
that of a {\it massless} relativistic particle moving in (3 or 2-dimensional)
Minkowski space
where, however, the momenta $\tp_A$ of the particle --which are our new
configuration variables-- are subject to the non-holonomic constraint
$\tilde{p}_0\ge 0$ and, in some models, also $\tilde{p}_+ >0$. Inspite of
these constraints, in each of these models the \ps\ (as well as the reduced
phase space) still has the structure of a cotangent bundle over the new
configuration space, spanned by the $\tp_A$.

\section{Quantization}

This section is divided into three parts. In the first, we outline the
general strategy, in the second, we carry out the quantization program
systematically in the case of the type II model, and, in the third we
briefly discuss the type I model. Since these two models are the
prototypes for the rest, other DIMT models can be treated in a completely
analogous fashion; they will not be discussed further.

\subsection{Outline}

We now want to exploit the simplicity of the scalar constraint in the
tilde canonical variables to carry out quantization.

There exist two standard procedures for quantization of systems with
first class constraints: the reduced phase space method and the Dirac
procedure of imposing operator constraints to select the physical
states. In all the models considered in this paper, if we use $\tp_A$
as the configuration variables, the constraints are independent of momenta.
Hence, quantization via the reduced phase space method leads to the same
result as quantization via Dirac's operator constraint method. Since our
primary interest stems from the quantization program of
\cite{newbook,aa:rst} which is an extension of Dirac's procedure,
we will use the operator constraint method.

In broad terms, the DIMT models fall into two classes: those in which
the effective configuration space is the full future null cone \cone\
in 2 or 3 dimensional Minkowski space and those in which it is only
the right half \halfcone\ of this null cone.
The mathematical structure of the models in the first class is
identical to that of a free relativistic particle in 2 or 3 dimensional
Minkowski space. Therefore, as far as the mathematical steps in the
quantization program are concerned, one can simply mimic the textbook
procedure for quantization of the relativistic particle. In models in
the second class, however, certain subtle issues arise because
of the presence of the non-holonomic constraints $\tp_{+} >0$. Therefore,
we will first treat this case in detail and then turn briefly to models
in the first class. For concreteness, we will use the type II and type I
models as representatives of the two classes.

\goodbreak
\subsection{The type II model}

Let us follow the quantization program of
\cite{newbook,aa:rst} step by step to bring out the
assumptions involved. Since the model is simple enough, the final
result is not surprising. However, in order to compare and contrast
this result with the one obtained in section 6, it is important to
note the procedure carefully. Also, in several other cases, treated
elsewhere \cite{aa:rst,rst:thesis}, the final result is not at all
obvious from the start and can in fact be quite surprising. In such
cases, it becomes critical to adhere to the program systematically.

To begin with, as in the Dirac approach to quantization of constrained
systems, let us ignore the scalar constraint. The phase space is then
topologically $R^6$ coordinatized by $(\tb^A, \tp_A)$, where $\tp_+$ and
$\tpz$ range over $(0, \infty)$ and all other coordinates range over
$(-\infty, \infty)$. It is natural to choose the $\tp_A$ as the configuration
variables and the $\tb^A$ as the conjugate momenta. Let us denote the
configuration space by ${\cal C}$. In the passage to quantum theory, let
us first consider the topological vector space ${\cal V}$ spanned by
distributions $\Psi$ over the space ${\cal C}$. This will be the initial
space of quantum states, prior to the imposition of constraints. Given
a smooth function $f(\tp)$ on ${\cal C}$ we define a configuration
operator $\hat{f}$ on ${\cal V}$ and given a complete, smooth vector
field $v$ on ${\cal C}$, we define a momentum operator $\hat{v}$ on
${\cal V}$ as follows:
\be \label{eq:repn}
 (\hat{f}\circ\Psi)(\tp) := f(\tp)\cdot\Psi(\tp) \qquad
 (\hat{v}\circ\Psi)(\tp) := i\hbar(\Lie_{v}+\fr12 ({\rm Div}_{\tilde\mu} v))
 \Psi(\tp )\ .
\ee
Here, $\tilde{\mu}$ is an arbitrary volume element (i.e., a
nonvanishing 3-form) on ${\cal C}$ (to be fixed later) and the
divergence of the vector field $v$ with respect to $\tilde\mu$ is
defined by: $(\Div{\tilde\mu} v)\cdot\tilde\mu :=\Lie_{v}\tilde{\mu}$. The
operators are defined in such a way that their commutators are
precisely $i\hbar$ times the Poisson brackets between their classical
analogs. Note that ${\cal V}$ is {\it not} equipped with the structure
of a Hilbert space. One can, if one is so inclined, introduce an inner
product and make ${\cal V}$ into a Hilbert space.  However, typically,
most solutions to constraints are not
normalizable with respect to such an inner product and the resulting
Hilbert space structure has no physical significance.

Our next task is to solve the quantum constraint equations
$\hat{C}_T\circ\Psi=0$, thereby singling out the physical states.
Since the scalar constraint function,
\be  \label{eq:calvin}
   C_T=\fr12(-\tpz^2 +\tpp^2 +\tpm^2)\ ,
\ee
is a smooth function of $\tp_A$ alone, this task is easy to accomplish.
Physical states lie in the vector space of solutions to the quantum
constraint equation, which is spanned by states of the form:
\be  \label{eq:ps}
 \Psi_{\rm sol}(\tp) = \delta(\tp_0 -\sqrt{\tp_+^2 +\tp_-^2})\cdot
 \psi(\tp)\ ,
\ee
where (as before) $\Psi_{\rm sol}$ is a 3-dimensional distribution on
${\cal C}$ and $\psi$ is a 2-dimensional distribution on \halfcone.
(That is, we have to smear $\Psi_{\rm sol}$ by a test field on the
3-dimensional space ${\cal C}$ while we have to smear $\psi$ by a test
field on the right half of the 2-dimensional future light cone
\halfcone\ in ${\cal C}$.  Note that since $\tpz\in(0,\infty)$, the
non-holonomic constraints are also satisfied.) Denote the space of
these solutions by ${\cal V}_{\rm sol}$. Since the distribution
$\delta(\tp_0 -
\sqrt{\tp_+^2+\tp_-^2})$ is a pre-factor common to {\it all} solutions,
each state $\Psi_{\rm sol}$ is completely characterized by the distribution
$\psi$ on \halfcone .

We can now consider physical --i.e., Dirac-- observables. These are
the operators which leave the space ${\cal V}_{\rm sol}$ of solutions
invariant. Every $\hat f$ defined above has this property.
However, since \halfcone\ is only 2-dimensional, two (suitably chosen)
operators among these suffice to constitute a complete set of physical
configuration operators. Let us choose these to be $\hat{\tilde{p}}_\pm$,
operator analogs of $\tp_\pm$.

The choice of a minimal, complete set of momentum operators requires
more care.  Now, the corresponding vector fields $v$ need to satisfy
four additional properties: $i)$ they must be tangential to \halfcone\ ;
$ii)$ they should span the tangent space to \halfcone\ ; $iii)$ they
should be closed under the Lie bracket; and $iv)$ the diffeomorphisms
they generate should leave invariant the three dimensional vector
space spanned by the two functions $\tp_\pm$ and constants. The first
two of these conditions ensure that the corresponding operators
$\hat{v}$ are well-defined and form a complete set of momentum
operators while the last two conditions ensure that the vector space
generated by the configuration and momentum operators (together with
the identity) is closed under the commutator bracket.  Consider
the vector fields:
\be \label{eq:dvf} \eqalign{
   \tilde{v}_-&= \left(\frac\d{\d\tpm}\right)+\frac\tpm\tpz\left(
            \frac\d{\d\tpz}\right) \cr
    \tilde{v}_+&= \tpp\left(\frac\d{\d\tpp}\right) + \frac{\tpp^2}\tpz \left(
            \frac\d{\d\tpz}\right). \cr}
\ee
These vector fields span the tangent space at each point of \halfcone\
and are complete; in particular, on the ``boundary'' $\tpp =0$ (which is
{\it not} a part of \halfcone ), the coefficient of $\d/\d\tpp$ vanishes.
Let us denote the corresponding complete set of momentum
observables by $\hat{\tilde{v}}_-$ and $\hat{\tilde{v}}_+$. Their action on
solutions to the quantum constraint is given by:
\be\label{b2:momops}
 (\hat{\tilde{v}}_\pm\circ \Psi_{\rm sol})(\tp) = i\hbar\delta(\tp_0
 -\sqrt{\tp_+^2 +\tp_-^2})(\Lie_{\tilde{v}_\pm} +\fr12 (\Div{\tilde\mu}
 \tilde{v}_\pm))  \psi(\tp)\ .
\ee
Since $\tilde{v}_\pm$ are tangential to \halfcone\  the action maps physical
states to physical states. Next, a straightforward calculation shows that
the vector fields $\tilde{v}_\pm$ --and hence also the corresponding operators
$\hat{\tilde{v}}_\pm$-- commute with one another. Finally, the nonvanishing
commutators between the configuration and the momentum operators are
given by:
\be \label{eq:alg}
   \eqalign{
   [\hat{\tilde{v}}_-, \hat{\tilde{p}}_- ] &= i\hbar \quad \hbox{and} \cr
   [\hat{\tilde{v}}_+, \hat{\tilde{p}}_+] &=i\hbar \hat{\tilde{p}}_+\ .\cr
}\ee
Thus, our choice of vector fields $\tilde{v}_\pm$ satisfies all the four
conditions required above. The complete set of Dirac observables is therefore
given by $\hat{\tilde{p}}_\pm$ and $\hat{\tilde{v}}_\pm$ and they provide
us with two ``canonically conjugate'' pairs.

We now come to the problem of finding an inner product. It is here
that we bring in the ``reality conditions'': The inner product should
be such that the Dirac operators corresponding to {\it real} classical
observables should be self-adjoint \cite{newbook,aa:rst}. Since the above set
of four Dirac observables (together with the identity operator) is
complete, i.e.\ since it generates the entire algebra of physical
observables, it suffices to impose the reality conditions only on this
set. Consequently, it follows from the results of \cite{rendall:pc} that,
if there exists an inner product on an irreducible representation of the
algebra of Dirac observables which makes these four Dirac observables
self-adjoint, the inner product is unique. To show existence, it will
suffice to simply exhibit the answer. For any two solutions $\Psi_{\rm sol}$
and $\Phi_{\rm sol}$ (see (\ref{eq:ps})), the required inner product is
given by:
\be\label{b2:ip}
  {\IP\Psi\Phi}_{\rm phy}\equiv \IP\psi\phi
       =\int_{\scriptstyle\halfcone}\> \bar{\psi}\, \phi\,\mu\ .
\ee
Here $\mu$ is the measure on \halfcone\ (i.e.\ a nonvanishing 2-form) such
that
\be\label{hobbes}
 \mu\wedge\rd C_T=\tilde{\mu},
\ee
where $\tilde{\mu}$ is the measure on ${\cal C}$ used in the
expressions of the momentum operators (\ref{b2:momops}), and $\rd C_T
$ is a nowhere vanishing covariant normal to \halfcone. Now, for any
vector field $v$ on ${\cal C}$, tangential to \halfcone, it is
straightforward to show that $\Div{\tilde\mu}v=\Div{\mu} v$, where the
(2-dimensional) divergence of $v$ on \halfcone\ is defined by
$(\Div{\mu} v)\cdot \mu :=\Lie_v\mu$. Thus, it follows that the
physical momentum operators (\ref{b2:momops}) are self-adjoint with
respect to the inner product (\ref{b2:ip}). (Since we are working in a
representation in which the operators $\hat{\tilde{p}}_\pm$ are
diagonal, these operators are self adjoint for {\it any} choice of
measure on \halfcone; the constraint on the measure comes only from
the requirement that the momentum operators $\hat{\tilde{v}}_\pm$ be
self adjoint.)  The physical states are those $\Psi_{\rm sol}$ which
have finite norm. This condition selects, from all distributional
solutions $\Psi_{\rm sol}$ to the quantum constraints, the ones for
which $\psi$ are square integrable functions on \halfcone\ (w.r.t.\ the
measure $\mu$). The Hilbert space $\tilde{\cal H}$ of physical states
is simply $L^2(\halfcone, \mu)$.

Recall that there is considerable freedom in the choice of $\mu$; it
can be {\it any} nowhere vanishing measure on \halfcone . Different
choices of $\mu$ lead to unitarily equivalent theories: If we were to
replace $\mu$ by $\mu' = f^2\mu$, then the map $\psi\mapsto\psi' =
\psi/f$ is the required unitary mapping from $\tilde{\cal H}$ to $\tilde{{\cal
H}'}$. One can use this freedom to simplify the expression
(\ref{b2:momops}) of the momentum operators $\hat{\tilde{v}}_\pm$. The
simplest expression results if we choose the measure $\mu$ with
respect to which the vector fields $\tilde{v}_\pm$ are
divergence-free. This condition provides a set of differential
equations on $\mu$ whose solution yields the expression
\be
  \mu_0=\frac1\tpp\> \rd\tpp\wedge\rd\tpm \ ,
\ee
For simplicity, from now on, we shall use this measure.

This completes the implementation of the quantization program for the
Bianchi type II model. In the final picture, the Hilbert space
$\tilde{\cal H}_0$ of physical quantum states is given by $L^2(\halfcone
, \mu_0)$; a complete set of Dirac observables is given by
$\hat{\tilde{p}}_\pm$ and $\hat{\tilde{v}}_\pm$; and, the algebra of
Dirac operators is given by the relations (\ref{eq:alg}).  This simple
picture has emerged precisely because we could cast the scalar
constraint function in the simple form (\ref{eq:calvin}) via a
canonical transformation to the $(\tb^A ,\tp_A)$ variables. The issue
of physical interpretation of this mathematical framework will be
discussed in detail in section 5.

\subsection{The type I model}

In this subsection, we wish to consider models in which the constraint
allows $\tilde{p}_A$ to belong to the {\it full} (future) null cone.
This discussion will also shed light on some subtleties that underlie the
choices of Dirac observables made above for the type II model.

The prototype of the models with the full light cone is provided by Bianchi
type I. Let us restrict ourselves to this case. Then, the constraint in the
original $(\beta^A, p_A)$ variables is already in the desired form since
there is no potential term to begin with. Therefore, there is no need to
carry out any canonical transformation; the tilde variables are the
same as the original ones. To emphasize this point and also to
distinguish this case notationally from the type II model, we will
work with the original phase space coordinates $(\beta^A, p_A)$, all
of which range over $(-\infty,\infty)$.

We can again follow the quantization program step by step. The first
few steps are identical to those for Bianchi type II: On the vector space
of distributions $\Psi$ on the new configuration space, we have a
representation of the configuration and momentum operators as in
(\ref{eq:repn}) and the quantum constraint for the type I model is
solved by states analogous to those in (\ref{eq:ps}). The only technical
difference is that there is only one non-holonomic constraint, namely the
one which restricts $p_0$ to be positive, whence the 2-dimensional
distributions $\psi$ which characterize the solutions to the quantum
constraint are defined on the {\it entire} future light cone \cone.
A complete set of classical Dirac configuration observables is given by
$p_\pm$. We will denote the corresponding quantum operators by $\hat{p}_\pm$.

However, for the momentum observables, there is a key difference from
the type II model: Operators analogous to $\hat{\tilde{v}}_\pm$ will no
longer suffice because now the vector fields $\tilde{v}_\pm$
span the tangent space of \cone only almost everywhere; they are linearly
{\it dependent} on the lines $\pp =0$. Hence we must choose different
Dirac momentum observables. Perhaps the simplest choice is to use
the two boost generators:
\be \label{eq:Iboosts}\eqalign{
  v_+&=\pz\left(\frac\d{\d\pp}\right) + \pp\left(\frac\d{\d\pz}\right)\cr
  v_-&=\pz\left(\frac\d{\d\pmm}\right) +
     \pmm\left(\frac\d{\d\pz}\right)\ , \cr}
\ee
which {\it are} linearly independent everywhere on (and tangential to)
\cone . However, this set fails to be closed under the Lie bracket: the
bracket of two boosts is a rotation. Thus, we must enlarge the set of
momentum observables by adding the rotation vector fields
\be
 v_0 =\pp\left(\frac\d{\d\pmm}\right) - \pmm\left(\frac\d{\d\pp}\right)\ .
\ee
Denote the momentum operators by $\hat{v}_A$. Now, if we compute the
commutators between these momentum operators
and the configuration ones, we find that the Lie algebra does not close
unless we add to the configuration operators $\hat{p}_0 \equiv
((\hat{p}_+)^2 +(\hat{p}_-)^2)^{\half}$. These six Dirac operators,
$(\hat{p}_A, \hat{v}_A)$ form a Lie algebra which is isomorphic to the Lie
algebra of the Poincar\'e group in 3-dimensional Minkowski space: the
configuration operators provide the generators of translations while the
momentum operators provide the generators of Lorentz transformations.%
\footnote{However, since here the effective configuration space \cone\ is
 only 2-dimensional, there are two algebraic relations between the
 above six operators. We could have avoided this redundancy by choosing the
 Dirac observables differently: For example, we could have chosen momentum
 operators corresponding to the vector fields
\begin{eqnarray*}
  u_+&=&\left(\frac\d{\d\pp}\right) + \frac\pp\pz\left(\frac\d{\d\pz}\right)\\
  u_-&=&\left(\frac\d{\d\pmm}\right) +
     \frac\pmm\pz\left(\frac\d{\d\pz}\right)\ .
\end{eqnarray*}
 These vector fields commute amongst themselves, and the Dirac
 operators $(\hat{u}_\pm,\hat{p}_\pm)$ form two canonically conjugate
 pairs.  With this choice, however, the quantum description would not have
 closely resembled the textbook treatment of a free relativistic
 particle.}
The classical Dirac observables corresponding to these six operators
are real, whence the six operators have to be self adjoint in the
quantum theory. Given the explicit expression of the momentum
operators (\ref{eq:repn}), which involves the choice of a measure on
\cone , the above ``reality condition'' again selects the inner
product (of the form (\ref{b2:ip})) uniquely. If the measure is chosen
so that all three vector fields are divergence-free (to simplify the
expressions of the momentum operators) we are led to the measure
$\mu'_0=(1/\pz)\rd\pp\wedge\rd\pmm$. Thus, not surprisingly, the
program has led us to a description which is the same as the textbook
treatment of the free relativistic particle in 3-dimensional Minkowski
space \cite{newbook,aa:rst}. The Hilbert space is now the space
$L^2(\cone , \mu'_0)$, the space of square-integrable
functions on the {\it entire} future cone \cone\ in the configuration
space spanned by $p_A$, where the measure $\mu'_0$ is given
above.

What would have happened if we had ignored the fact that $\tilde{v}_\pm$ are
not linearly independent {\it everywhere} and used the same set of
Dirac observables as in the case of the Bianchi II model? Then, the
Hilbert space $L^2(\cone ,\mu'_0)$ constructed here would have
provided a {\it reducible} representation of the algebra of those
Dirac observables. Since, for physical reasons, one must restrict
oneself to an {\it irreducible} representation, we would have been led
to use, as the space of physical states only ``half of'' $L^2(\cone
,\mu'_0)$, which would clearly have been wrong. Thus, strict
completeness of Dirac observables is important for quantization.
Considerable care must be exercised even in the case when completeness
fails on sets of measure zero%
\footnote{This point is important to quantization of the full 2+1 as well
as 3+1 dimensional general relativity using loop variables since these fail
to be complete on sets of measure zero on the classical phase space
(see, e.g., \cite{newbook,aa:rev}). In the 2+1 theory, one knows how
to face the resulting difficulties. In the 3+1 theory, however, the issue
is only partially understood.}.

Similarly, in the case of the type II model, it would be wrong to
ignore the global structure of the configuration space --i.e., the
non-holonomic constraint $\tpp > 0$-- and use the generators of boosts
and rotations as Dirac observables. That algebra would have led us to
the Hilbert space $L^2(\cone,\mu'_0)$ on the full future cone
rather than $L^2(\halfcone ,\mu_0)$ obtained in the previous
sub-section. Indeed, this difference in the global structure is the
only remnant in the tilde variables $(\tb^A, \tp_A)$ of the potential
term $U_T$ in the scalar constraint in the original ADM variables
$(\beta^A, p_A)$. It is therefore crucial to keep track of it in the
quantization procedure.

\goodbreak
\section{Physical Interpretation}

This section is divided into 3 parts. In the first, we state the
problem we wish to address and outline the general strategy (see e.g.\
\cite{aa:rst}).  This is implemented in detail for the simplest DIMT
model --Bianchi type I-- in the second part and for the the prototype
of the remaining DIMT models --Bianchi type II-- in the third part.
The overall situation in other DIMT models is analogous to that in
these two cases.

\subsection{The problem}

In section 4, we carried out the quantization program of
\cite{newbook,aa:rst}
to completion for all DIMT models. Since we were able to construct
complete sets of Dirac observables, we could use the ``reality
conditions'' to select the appropriate Hilbert space structures on the
spaces of physical states. Contrary to a general belief (see, e.g.,
\cite{kk:qg2}), we did not have to single out time and deparametrize the
system in order to arrive at this mathematical description.

Since we have access to a complete set of Dirac observables, we can pose
and answer a number of physical questions. For example, we can compute
the spectra of these observables; comment on their continuous versus
discrete eigenvalues; evaluate their expectation values in given physical
states, thereby providing probabilistic estimates for finding any given
range of values on any given state; etc. These {\it are} interesting
questions. However, of necessity, they all refer only to Dirac observables:
the action of more general operators fails to be well-defined since they do
not even leave the Hilbert space of physical states invariant.

Now, in the classical theory, dynamics is governed by the scalar constraint
whence Dirac observables are, in particular, constants of motion. We
therefore expect that, in quantum theory as well, questions which are
formulated using {\it only} the Dirac observables introduced so far
--$(\hat{\tilde{p}}_\pm , \hat{\tilde{v}}_\pm)$ in the type II model and
$(\hat{p}_A, \hat{v}_A)$ in the type
I model-- will also refer to physical quantities which do not ``evolve.''
After all, we have a framework in which there is no notion of time and hence
of evolution. Nothing ``happens.'' So far, there is only a timeless, frozen
formalism. However, we would like to ask questions, e.g., about evolution of
anisotropies, about the behavior of spacetime curvature, about the fate of
classical singularities in the quantum theory. The machinery of Dirac
observables at hand does not in itself suffice to even phrase such questions.
Neither the anisotropies $\beta^A$ nor the curvature scalars such as
$C_{abcd}C^{abcd}$ commute with the constraint; they are {\it not}\
expressible purely in terms of our Dirac observables. Thus, there seems to
exist a quandary: while the mathematical machinery wants us to work primarily
in terms of Dirac operators, many of the interesting physical questions refer
to ``dynamics'' and hence, on the face of it, seem to force us beyond Dirac
observables.

There is, however, a well defined strategy that one can adopt to get
out of this apparent quandary. The idea is to isolate, prior to the
imposition of the constraint, one of the arguments of the physical
wave functions as an ``internal time variable'' and interpret the
constraint as an evolution equation with respect to this internal
time.%
\footnote{For a review of other approaches to the problem of time in
quantum gravity see \cite{kk:cji}.}
One can then introduce new {\it one parameter families}\ of Dirac
operators and interpret the non-trivial dependence on the parameter as
``time evolution.'' Note, however, that this ``time'' is one of the
configuration variables. It does {\it not} arise from a background
space-time; at a fundamental level, there is in fact no space-time
whatsoever in the quantum theory. Nonetheless this generalized notion
of time appears to be sufficient to pose and answer the dynamical
questions raised above.  Furthermore, it appears to suffice also for
the analysis of measurement theory since, as we will see below, one
can specify exhaustive sets of mutually exclusive alternatives on
slices of constant (generalized) time in the configuration space. To
summarize, although it may seem puzzling at first, it {\it is}
possible to introduce ``time evolving Dirac observables'' in
an appropriate sense and use them effectively to analyse the questions of
dynamics. It is this strategy that lets us get out of the apparent quandary.

This general strategy is of course rather old (see, e.g., articles by
Kucha\v r and Rovelli in \cite{as:osgood} and the references they contain)
although its full power does not seem to be always appreciated. What we
wish to show in the next two subsections is: $i)$ this strategy can be
implemented in detail in all DIMT models; and, $ii)$ the implementation
enables us to ask and answer a number of ``dynamical'' questions of physical
interest, including the fate of singularities. This goes a long way towards
understanding the physics in the quantum theory of these spatially homogeneous
models. Our overall conclusion is that while it is {\it not} essential to face
``the issue of time'' to complete the quantization program itself, a
satisfactory treatment of this issue is necessary to extract the {\it full}
physical content of the resulting mathematical framework and that this can
be achieved for all DIMT models.

\goodbreak
\subsection{The type I model}

Let us begin by noting that not all mathematically equivalent representations
in quantum theory are suitable for addressing the issue of time. (For
details, see, e.g., the article by Ashtekar in \cite{as:osgood}.) For the
free relativistic particle, for example, time is explicit in the position
representation; the quantum constraint equation, $\eta^{ab}\nabla_a\nabla_b
\Phi(x) = 0$ can be immediately interpreted as the evolution equation. In
the momentum representation, by contrast, the constraint equation
$\eta^{ab}\hat{p}_a\hat{p}_b \circ\Phi(p) = 0$, does not have the form of
an evolution equation at all. Not surprisingly, the situation is the same
in DIMT models. This is why the use of the momentum representation led to
a frozen formalism in section 4.

In the type I model, the minimum change necessary is to consider, in the
very beginning, wave functions which depend not on $p_A$ but rather on
$(p_\pm, \beta^0)$. (Alternatively, we can use the three $\beta^A$ as
arguments of the wave functions; the essential point is only that the
representation be diagonal in $\beta^0$.) Then, the quantum constraint
equation becomes:
\be \label{eq:qc}
 -i\hbar \d_0\Phi (p_\pm, \beta^0 ) = \sqrt{\hat{p}_+^2 +\hat{p}_-^2}
 \cdot \Phi(p_\pm ,\beta^0) \ ,
\ee
where $\d_0=\d/\d\beta^0$. Note that we have also incorporated the
non-holonomic constraint $p_0 \ge 0$.  (See also footnote 1.) This
equation is easy to integrate:
\be \label{eq:sol}
 \Phi(p_\pm ,\beta^0) = (\exp ({i\over\hbar} \sqrt{p_+^2 +p_-^2}\>\beta^0))
 \cdot \phi(p_\pm)
\ee
Denote as before the space of these states by ${\cal V}_{\rm sol}$. Physical
states will be normalizable elements of ${\cal V}_{\rm sol}$.

The six operators $(\hat{p}_A , \hat{v}_A )$ of section 4.3 continue to
provide a complete set of Dirac observables. On the space of solutions
(\ref{eq:sol}), their explicit expressions reduce to: $\hat{p}_\pm \circ\phi
= p_\pm\cdot \phi$;  $\hat{p}_0 \circ \phi = \sqrt{p_+^2 + p_-^2}\cdot \phi$;
$\hat{v}_\pm\circ\phi = i\hbar\sqrt{p_+^2 +p_-^2}\cdot \d_\pm \phi$;  and,
$\hat{v}_0\circ\phi = i\hbar (p_+\d_- - p_-\d_+)\phi$, where
$\d_\pm=\d/\d p_\pm$. Once again, we can
use the ``reality conditions'' to arrive at the inner product. We begin with
a general measure $\tilde\mu(p_\pm, \beta^0)$ on the domain space of the
solutions $\Phi(p_\pm, \bz)$ to (\ref{eq:qc}), write the inner product as
\be
 \IP{\Psi(p_\pm, \bz)}{\Phi(p_\pm ,\bz )} = \int \ \tilde\mu(p_\pm, \beta^0)\,
 d\beta^0\wedge dp_+\wedge dp_- \ \overline{\Psi(p_\pm , \beta^0)}
 \Phi(p_\pm, \beta^0)\ ,
\ee
and constrain the measure by requiring that the six Dirac observables
be self-adjoint with respect to this inner product. This condition
restricts the measure to have the form $\tilde\mu(p_\pm, \beta^0) = \mu(\bz)
/\sqrt{p_+^2 +p_-^2}$; the dependence on $p_\pm$ is completely
determined while that on $\beta^0$ is left unconstrained. Thus, the
inner-product compatible with the reality conditions must have the form:
\be \label{eq:IPI}
 \eqalign{
 \IP{\Psi(p_\pm, \bz)}{\Phi(p_\pm ,\bz )} =& \int\
 {\mu(\bz)\over\sqrt{p_+^2+p_-^2}} (\rd\bz \wedge \rd p_+\wedge
 \rd p_-)\ \bar\Psi(p_\pm ,\bz ) \ \Phi(p_\pm ,\bz )\cr
 =& K \int_{\bz ={\rm const}} {{\rd p_+\wedge \rd p_-}\over
 \sqrt{p_+^2+p_-^2}}\ \bar\Psi(p_\pm ,\bz )\ \Phi(p_\pm ,\bz )\cr
=& K \int_{\bz ={\rm const}} {{\rd p_+\wedge \rd p_-}\over
 \sqrt{p_+^2+p_-^2}}\ \bar\psi(p_\pm )\ \phi(p_\pm ) \ , \cr}
\ee
where we have used the constraint equation (\ref{eq:qc}) in the
passage to the second step and where $K=\int d\bz \mu(\bz )$ is the
total measure of the $\bz$ line with respect to $\mu(\bz )$.  Note
that because of the constraint equation, the final integral is
independent of the value of $\bz$ at which the integral is evaluated
and has the same form as in the frozen formalism of section 4.
However, we did not {\it have} to deparametrize the theory (in the sense of
\cite{kk:qg2}) and slice the configuration space with $\bz ={\rm const}$.
slices in order to {\it arrive at} these expressions of the inner product.
We began with a {\it general} measure on the 3-dimensional domain space
and used the reality conditions on Dirac observables {\it to conclude}
that the inner product must have the form given above. For simplicity, from
now on we will assume that $\mu(\bz)$ is chosen so that the constant $K$
is normalized to unity.

Our next task is to introduce ``time dependent Dirac observables.''
Let us begin with the anisotropies $\beta^A$. Clearly, as they stand,
$\beta^A$ themselves are {\it not} Dirac observables since they do not
commute (even weakly) with the constraint; if $\Phi$ is a physical state,
$\hat\beta^A\circ\Phi$ is not. However, using the fact that the
physical states satisfy (\ref{eq:qc}), we can construct a {\it
one-parameter family of Dirac operators} $\hat\beta^A(t)$,
parametrized by a real parameter $t$.  To see this, let us begin by
noting that, because of (\ref{eq:qc}), every physical state
$\Phi(p_\pm ,\bz)$ is completely determined by its value on a $\bz =
const$ surface.  Hence, we can define the operators $\hat\beta^A(t)$
as follows: To act on a physical state $\Phi(p_\pm ,\bz )$,
freeze it on the surface $\bz = t$, act on it by the operators
$\hat\beta^A$ and evolve the resulting function to all values of $\bz$
using the constraint equation (\ref{eq:qc}). (Here, the action of $\hat
\beta^A$ on the ``frozen'' states is the obvious one: $\hat\bz\circ
\Phi(p_\pm,\bz =t) = t\cdot\Phi(p_\pm, \bz=t)$ and $\hat\beta^\pm\circ
\Phi(p_\pm, \bz =t) = i\hbar\d_\pm\Phi(p_\pm, \bz=t$). By construction,
the state $\hat\beta^A(t)\circ\Phi$ satisfies (\ref{eq:qc}) and is thus
again a physical state. The explicit expression of these operators is
given by:
\be \label{eq:tddo}
 \hat\beta^A(t)\circ\Phi(p_\pm, \bz) := e^{i\hat{H}(\bz - t)}\circ
 \hat\beta^A\circ \Phi(p_\pm,\bz = t)\ ,
\ee
where we have set $\hat{H} =(1/\hbar)\sqrt{(\hat{P}_+^2
+\hat{P}_-^2)}$. It is straightforward to check explicitly that
{\it for each real number} $t$, the operators $\hat\beta^A(t)$ commute
weakly with the scalar constraint. (For details, see e.g.\ the analogous
construction for the nonrelativistic parametrized particle in \cite[\S 6.2]
{rst:thesis}.)

Note that $\hat\bz(t)$ is just a multiple of identity, $\hat\bz(t)\circ\Phi
= t\Phi$, and therefore commutes with every other operator. In each
classical solution, $\bz$ increases monotonically in time and can be
taken to be an internal time parameter. The expression of $\hat\bz(t)$
therefore suggests that it is natural to interpret the parameter $t$ as a
generalized time in quantum theory. The parameter $t$ of course does not
refer to any
specific space-time, whence the adjective ``generalized.'' However, on
semi-classical states an approximate space-time interpretation is
possible. Note finally that in the classical theory, the volume $V$ of the
spatially homogeneous slice is given by $V = \exp{3\bz}$. Therefore, in
quantum theory, the 1-parameter family of Dirac observables $\hat{V}(t) :=
\exp 3\hat\bz(t)$ represents the ``volume operator at time $t$.'' Its action
on physical states is $\hat{V}(t)\circ \Phi = \exp{(3t)}\cdot \Phi$.
Thus, the volume observable increases monotonically in time also on
the quantum physical sector.

As in the classical theory, the anisotropies $\hat\beta^\pm (t)$ are
the two genuine dynamical quantities. Given any two physical states
$\Phi,\Psi$, one can study the dependence on the parameter $t$ of the
transition amplitudes $\IP{\Phi} {\hat\beta^\pm(t)\circ\Psi}$. In particular,
for $\Phi=\Psi$ this tells us how the expected quantum anisotropies
evolve in that state. One can similarly introduce the time dependent Dirac
observables $\hat{p}_A (t)$. However, since $\hat{p}_\pm$ commute with
$\hat{H}$, $\hat{p}_\pm(t)$ turn out not to depend on the parameter value $t$.
Furthermore, from (\ref{eq:qc}) it follows that $\hat{p}_0 =(\hat{p_+}^2 +
\hat{p_-}^2)^\half$. Thus, as far as the momentum operators are concerned,
the procedure of (\ref{eq:tddo}) just leads us back, as one might expect,
to the Dirac observables introduced in section 4.3. Given any instant $t_0$
of time, we now have a complete set of Dirac observables $(\hat\beta_\pm(t_0),
\hat{p}_\pm(t_0)$. These satisfy the canonical commutation relations. However,
since $\hat\beta^0(t_0)= t_0\hat{1}$ is just a multiple of identity and
$\hat{p}_0(t)$ is algebraically related to $\hat{p}_\pm$, the commutators
involving $\hat\beta^0(t_0)$ and $\hat{p}_0(t)$ do not mirror the corresponding
Poisson brackets on the unconstrained phase space. Finally, given a general
classical observable, $F(\beta^\pm, p_\pm)$,  one can
construct the corresponding one parameter families $\hat{F}(\beta^\pm,p_\pm)
(t)$ of Dirac operators and study their dependence on the (generalized) time
parameter $t$ as follows:
\be \label{eq:tddogen}
 \left(\hat{F}(\beta^A,p_A)(t)\circ\Phi\right)(\bz, p_\pm) :=
 e^{i\hat{H}(\bz - t)}\circ F(\hat{\beta}{}^A,\hat{p}_A)\circ
 \Phi(p_\pm, \beta^0=t)\ .
\ee
In practice of course one often encounters difficult factor ordering problems
in this procedure. Conceptually and technically, however, these are on the
same footing as the analogous problems encountered already in non-relativistic
quantum mechanics where there are no constraints and no problem of time.

Of particular interest to us is the Weyl curvature scalar $W:=C^{abcd}
C_{abcd}$. This quantity diverges at the singularity which occurs at $\bz
= -\infty$ in {\it all} non-flat classical solutions. What is the situation
in the quantum theory? Do these singularities persist or do they get washed
away due to ``quantum fuzzing''? Since the time evolution of (\ref{eq:sol})
is unitary with respect to the inner product (\ref{eq:IPI}), one might at
first expect that the quantum evolution is free of singularities. To see if
this is the case, let us construct the 1-parameter family of Dirac
observables $\hat{W}(t)$ and examine their dependence on the parameter $t$.
A simple calculation using (\ref{eq:tddogen}) yields:
\be\eqalign{
 \hat{W}(t)\circ\Phi(p_\pm,\bz)
  &= \fr{1}{3\cdot24^2} e^{i\hat{H}(\hat\bz -t)}\circ
 e^{-12\hat\bz}\hat{p}{}_0^4 (1+ \cos 3\hat\theta )\circ
 \Phi(p_\pm,\bz=t)\cr
 &= \fr{1}{3\cdot24^2}  e^{-12t}\hat{p}{}_0^4 (1+ \cos 3\hat\theta )
 \circ\Phi(p_\pm,\bz)\ ,\cr}
\ee
where $\tan\hat\theta = (\hat{p}_-/\hat{p}_+)$ and $\hat{p}_0^2 =
(\hat{p}_+^2 +\hat{p}_-^2)$. Thus, as $t$ tends to $-\infty$,
$\hat{W}(t)\circ\Phi$ diverges on {\it every} normalizable state%
\footnote{Since the set defined by $1+\cos 3\theta=0$ is of measure zero,
 states with support just on this set are either indistinguishable from
 zero or genuinely distributional and hence not normalizable. Note also
 that the classical solutions corresponding to initial data in this set
 are (locally) flat and hence non-singular; the issue of ``quantum
 fuzzing'' is therefore irrelevant in any case for this set.}.
Alternatively, it is easy to
check that $\hat{V}^4(t)\hat{W}(t)$ is a time {\it independent} Dirac
observable. (There is no factor ordering ambiguity in this product.)
{}From its definition, it follows trivially that as the parameter $t$
goes to $-\infty$, the operator $\hat{V}(t)$ goes to zero, whence the
Weyl scalar $\hat{W}(t)$ diverges. {\it Thus, the singularity
persists inspite of the unitarity of quantum evolution.} (Similar
results have been obtained by Husain \cite{vh} for the Gowdy models.)

This may seem surprising at first since the mathematics of the model
is the same as that of a free relativistic particle. In that case, the
quantum theory is well-defined; there are no singularities. How does
this difference arise? Note first that the apparent paradox exists
already in the {\it classical} Hamiltonian description of the two
systems; it is {\it not} quantum mechanical in origin. The difference
arises because the physical interpretation associated with various
mathematical symbols is different in the two cases. In particular, the
analogs of anisotropies of the type I model are the position
coordinates of the relativistic particle. Let us first consider the
classical theory. On a generic particle trajectory, the position
coordinates of the particle tend to $\pm\infty$ as time goes to
$-\infty$. This ``divergence'' of course does {\it not} signal a
physical pathology. Since the underlying mathematics is identical, in
a generic type I solution, the anisotropies $\beta^\pm$ also diverge
as $\beta^0$ tends to $-\infty$. This divergence, on the other hand,
{\it does} represent a physical pathology. The {\it spacetime}
geometry becomes singular whence test objects, for example, would be
torn to pieces. The situation in quantum theory is completely
analogous. The same mathematical results can have drastically
different consequences because of differences in the physical
interpretation.

Could the above result on singularities have been anticipated on general
grounds? There is, for example, a viewpoint \cite{jw:amsci} that there should
be a rule of ``unanimity'': If generic classical solutions of the theory are
singular, the singularity would persist in the quantum theory. This is
indeed what we have observed in the type I model above%
\footnote{See also \cite[\S 6.4]{rst:thesis}, where however the reduced space
quantum theory is used and the deparametrization is carried out classically.
This approach yields the correct result for simple systems such as the type I
model now under consideration. In more general situations, however, one must
use a genuinely quantum mechanical deparametrization, e.g., using
(\ref{eq:tddo}). }.
However, it seems difficult to arrive at such a conclusion on general grounds.
Consider, for example, a particle moving in an attractive Coulomb potential,
subject to the condition that its angular momentum be zero. One then has
radial motion and {\it every} classical solution is singular in the sense
that the potential energy, for example, diverges in a finite time interval
along any dynamical trajectory. The quantum theory, on the other hand is
well defined: it corresponds to the spherically symmetric sector of the
Hydrogen atom problem. In particular, there is a dense subspace of the Hilbert
space on which matrix elements of the potential term $1/r$ remain finite for
{\it all} times. In this sense, the quantum dynamics is {\it very} different
from classical dynamics.

To conclude, note that using the prescription of (\ref{eq:tddogen})
one can also construct a 1-parameter family of Dirac observables
starting from the original six Dirac observables $\hat{p}_A$ and
$\hat{v}_A$. We saw above that for the three $\hat{p}_A$, the
dependence on $t$ drops out. The same is true for the three
$\hat{v}_A$. Thus, what distinguishes these six from a generic time
dependent Dirac observable $\hat{F}(t)$ is that these six are time
{\it independent}. To obtain the inner product via reality
conditions, it suffices to work with a complete set of Dirac
observables which may be time independent. If one can find such a set,
deparametrization is {\it not} needed to find the inner product on the
physical states. However, even in this case, to extract the dynamical
content of the theory in the usual sense, we need access to generic
time dependent Dirac observables.

\goodbreak
\subsection{The type II model}

The general line of argument in the type II case is the same as the
one given above: One introduces wavefunctions $\Phi(\tbz, \tp_\pm)$;
writes out the constraint as an evolution equation of the form
(\ref{eq:qc}); shows that the reality conditions lead one to the same
inner product as in section 4.2; and, introduces the time-dependent
Dirac observables analogous to (\ref{eq:tddogen}). The only technical
difference is that $\tp_+$ is now restricted to take
on just positive values.

The one parameter family $\hat{\tilde\beta}{}^0(t)$ corresponding to the
classical variable $\tbz$ is again $t$ times the identity operator on physical
states. Since $\tbz$ can be interpreted as time in the classical
Hamiltonian description, the parameter $t$ in the expressions of Dirac
operators can be again interpreted as (generalized) time in the quantum theory.
What is the relation between this time and the spatial volume $V$? Note
that $V$ is still given by $\exp 3\beta^0$ where $\beta^0$ is the {\it
original}\ Misner variable. Since we performed a non-trivial canonical
transformation to arrive at the tilde variables, the phase space time
variable $\tbz$ (and hence the quantum time variable $t$) is no longer
simply related to the spatial volume. Indeed, the volume is a rather
complicated function of the tilde canonical variables,
\be\eqalign{
 V^2 &= \exp 6\bz = \exp (4\sqrt{3}\bar\beta^0 + 2\sqrt{3}\bar\beta^+) \cr
  &= \sqrt{6} (\exp 4\sqrt{3}\tbz) \frac{ \cosh(2\sqrt{3}\tbp)}{\tpp}
      \ ,\cr}
\ee
and depends in particular on the {\it momentum} variable $\tpp$ as
well.  Nonetheless, the qualitative behavior of $V^2$ is similar to
that in the type I model. To see this, note first that along classical
dynamical trajectories, we have: $ \tilde{p}_\pm = {\rm const}$, and
$\tb^\pm = b^\pm - \tp_\pm\tbz/\sqrt{\tpp{}^2 +\tpm{}^2}$, where the
constants $b_\pm$ vary from one trajectory to another. It therefore
follows that the volume is
again a monotonically increasing function of the ``time parameter''
$\tbz$. Furthermore, as $\tbz$ tends to $-\infty$, the volume goes to
zero. Finally, in quantum theory, it is straightforward to compute the
(time-dependent) 1-parameter family of Dirac observables $V^2(t)$:
\be
 \hat{V}^2(t)= \sqrt6 e^{4\sqrt3 t}\cdot
 e^{i\hat{\tilde{H}}(\hat{\tilde\beta}{}^0-t)}\circ
 \half\left(\frac1{\hat{\tilde{p}}_+}\cosh(2\sqrt3\hat{\tilde\beta}{}^+)
 + \cosh(2\sqrt3\hat{\tilde\beta}{}^+) \frac1{\hat{\tilde{p}}_+}\right)
 \circ e^{-i\hat{\tilde{H}}(\hat{\tilde\beta}{}^0-t)},
\ee
where we have chosen the symmetric factor ordering and both sides act
on states $\Phi(\tbz,\tp_\pm)$, and where, as is the case for Bianchi type
I, the Hamiltonian operator is the ``free particle'' Hamiltonian:
\be \label{eq:derkins}
 \hat{\tilde{H}}=\frac1\hbar\sqrt{\hat\tpp{}^2+\hat\tpm{}^2}.
\ee
While the expression of the resulting operators is complicated, they are all
well-defined. One finds again that, in the limit as $t$ goes to $-\infty$,
the volume operator tends to zero.

To discuss the issue of singularities, we can, as before, analyse the behavior
of the Weyl scalar $W = C_{abcd} C^{abcd}$. As with the volume, the explicit
expression is rather complicated:
\bea
W&=&\frac{e^{-8\sqrt3\tbz}}{6\cdot18\cdot24}
     \bigg\{ 6\tpp^4(3\tpp^2+2\tpm^2)\ch{-2} \cr
   & & \quad +18\tpz\tpp^5\ch{-3}\sh -3\tpp^4(53\tpp^2+12\tpm^2)\ch{-4} \cr
   & & \quad  +108\tpz\tpp^5\ch{-5}\sh +172\tpp^6\ch{-6} \bigg\} \cr
 &=& \frac{e^{-8\sqrt3\tbz}}{6\cdot18\cdot24}
      \sum_{i=1}^5 Q_i(\tpz,\tpp,\tpm) T_i(\tbp),
\eea
where $Q_i$ are (low order) polynomials of their arguments and the $T_i$ are
hyperbolic trignometric functions. Along classical trajectories, as
$\tbz$ goes to $-\infty$, $W$ diverges at least
as fast as $\exp (-4\sqrt{3}\tbz)$. This is the initial singularity in the
classical theory. In the quantum theory, one can again construct the
1-parameter family of Dirac observables $\hat{W}(t)$ by using
(\ref{eq:tddogen}) and choosing the symmetric factor-ordering:
\be
\hat{W}(t)=\frac{e^{-8\sqrt3 t}}{6\cdot18\cdot24}\cdot
    e^{i\hat{\tilde{H}}(\hat{\tilde\beta}{}^0-t)}\circ \half
   \sum_{i=1}^5 \left( Q_i(\hat{\tilde{p}}_A) \circ
           T_i(\hat{\tilde\beta}{}^+) + T_i(\hat{\tilde\beta}{}^+) \circ
       Q_i(\hat{\tilde{p}}_A) \right)\circ
     e^{-i\hat{\tilde{H}}(\hat{\tilde\beta}{}^0-t)}.
\ee
The resulting operator $\hat{W}(t)$ again has the property that $\hat{W}(t)
\circ\Phi$ diverges for all states in the physical Hilbert space as $t$ tends
to $-\infty$. (For explicit computations, it is convenient to use the
representation in which the three $\tilde\beta^A$ --rather than
($\tilde{p}_\pm, \tilde\beta^0)$-- are diagonal.) This can be seen also by
considering, as for Bianchi type I, the behaviour of the operator
corresponding to $V^4(t)W(t)$. Similar considerations hold for the
rest of the DIMT models.

Thus, the curvature singularity persists in quantum theory for all DIMT
models.

To conclude this discussion, we point out how this notion of time can
be used to construct a meaningful measurement theory. In the quantum
mechanics of closed systems, such as the ones we are considering, to
discuss issues such as measurements, conditions under which states
decohere and behave semi-classically, and to make concrete physical
predictions, one generally begins with the notion of exhaustive sets
of mutually exclusive alternatives \cite{jh:lh}. For the models under
consideration, to obtain such sets we can foliate the domain space of
physical states by constant $\bz$ (in type I and $\tbz$ in type II)
slices. On a slice $\bz=t_0$, a complete set of alternatives can then
be constructed, e.g., using any one complete commuting set of (time
dependent) Dirac observables corresponding to {\it that} instant of
time, $t=t_0$. We can construct histories of physical interest --e.g.,
states which at time $t_1$ have anisotropies (which essentially
represent the 3-metric) in a specified range and which at time $t_2$
have momenta (essentially the extrinsic curvatures) in another
specified range, etc. Note also that, unlike in the path integral method,
we are not tying ourselves down to a specific configuration space: We can
easily switch from one representation to another, because we have full
recourse to the Dirac transformation theory. Hence, in this approach, we
can incorporate a large number of histories which cannot even be considered
in the standard path integral approaches.

Note however that we chose, right in the beginning, a preferred
deparametrization which is suggested by the form the constraint
assumed after the canonical transformation. What if some one makes
another choice? Is there a generalized ``transformation theory''
associated with such changes? As far as we are aware, one does not
even know how to phrase this question precisely in full generality.
One {\it can} make simple changes.  For example, after the canonical
transformation to the tilde variables, the
constraint took on the same form as that encountered in the treatment
of a free relativistic particle in Minkowski space (with, in some
models, the non-holonomic constraint $\tpp >0$.) Therefore, we could
have made a ``Lorentz transformation'' in the $\tb^A$ space and used
another choice of time with respect to which the constraint would
again have been of the form (\ref{eq:qc}). It is easy to see that the
resulting description would have been equivalent to the one given here
(provided the appropriate non-holonomic constraint is again imposed in
the quantum theory). What, however, if one made a completely different
choice of time with respect to which the constraint is again of the
desired form? This is the question that is wide open and should become
a focus of discussion on the issue of time. The main problem is that
we have rather limited understanding of {\it all} the choices which
render the constraint in the desired form, whence it is difficult to
make even a precise conjecture relating the many resulting
quantum descriptions.

For the Bianchi type II model, however, the Hamiltonian framework of
section 3.1 (see (\ref{eq:solvc})) does present
us with another deparametrization which is {\it  non-trivially}\
related to the one used in this section. Therefore, at least in this
one case, we can raise the question of equivalence. This is the topic of
discussion of the next section.

\goodbreak
\section{Different Quantization Strategies: Comparison}

In this section, we will restrict ourselves to the type II model. In this
case, already in the original variables $(\bar\beta^A, \bar{p}_A)$, prior
to the canonical transformation, the scalar constraint admits a conditional
symmetry \cite{kk:jmp}. That is, the configuration space, spanned by the
three $\bar\beta^A$, admits a vector field --$\d/\d\bbz$-- which is a
time-like Killing field of the supermetric, along the integral curves of
which the potential term in the constraint is constant. In the first
subsection, we outline the quantum theory \cite{atu:I} that results directly
by using this symmetry to decompose solutions to the constraint into positive
and negative frequency parts. A priori, it is not clear that this quantum
theory is equivalent to that presented in section 4 since the canonical
transformation to the tilde variables mixes coordinates and momenta. (In
the language of geometric quantization, the two methods use different
polarizations already at the kinematic level, before the imposition of the
constraint.) Furthermore, the two approaches adopt quite different techniques
to single out the inner product on the space of quantum states. In the
second subsection, we show that the two descriptions are in fact equivalent
in an appropriate sense.

In each case, one can deparametrize the theory. On the classical {\it phase
space}, the deparametrizations are equivalent since the canonical
transformation leaves $(\bar\beta^0, \bar{p}_0)$ unaffected; $\bar\beta^0 =
\tilde\beta^0$. In the quantum theory, however, the deparametrizations
are {\it not} obviously equivalent since the {\it domain spaces} of
physical states are quite different in the two cases due to the
non-triviality of the canonical transformation. Nonetheless, we will see
that equivalence does hold in an appropriate sense.

The final result holds also for the type I model. However, in this case,
the equivalence is hardly surprising: the barred canonical variables are just
linear combinations of the unbarred (or the tilde) variables since the
canonical transformation is now trivial.

\goodbreak
\subsection{Conditional symmetries}

Recall from section 3 that, in the type II model, the scalar constraint
in the barred variables is given by:
\be
 \half\eta^{AB} \bar{p}_A \bar{p}_B + 6 \exp (-4\sqrt{3}\bbp) = 0.
\ee
Consequently, the time-like (with respect to $\eta^{AB}$) vector field
$\d/\d\bbz$ on the configuration space spanned by the three ${\bar\beta}{}^A$
defines a momentum variable
$\bar{p}_0$ in the phase space, which Poisson-commutes with the scalar
constraint. Hence, the canonical transformation generated by $\bpz$ is
a classical symmetry and $\d/\d\bbz$ is a conditional symmetry in
the sense of \cite{kk:jmp}. Therefore, we can forego the construction
of a complete set of Dirac observables and the rest of the steps in
the quantization program \cite{newbook,aa:rst} which we used in
section 4 and carry out quantization by an entirely different route
\cite{kk:qg2,am:qf,k:cmp}. The idea here is as follows. Consider the
vector space $\bar{\cal V}_{\rm sol}$ of solutions to the quantum
constraint
\be \label{eq:qc1}
 \eta^{AB}\bar\d_A\bar\d_B\Phi(\bar\beta) +12 \exp (-4\sqrt{3}\bbp)
 \cdot\Phi(\bar\beta) = 0.
\ee
To equip $\bar{\cal V}_{\rm sol}$ with an appropriate Hilbert space structure,
we can use the conditional symmetry. The 1-parameter group of diffeomorphisms
generated by $\d/\d\bbz$ has a well-defined action on the space of solutions
to (\ref{eq:qc1}). We can therefore seek a Hermitian inner product on
$\bar{\cal V}_{\rm sol}$ such that this action is unitary; i.e., such that the
classical symmetry is promoted to the quantum theory.

The final result \cite{atu:I} can be summarized as follows. The Hilbert
space consists of normalizable solutions to the positive frequency part of
the quantum constraint (\ref{eq:qc1}), i.e., to the equation:
\be \label{eq:spiff}\eqalign{
 -i\hbar\bar\d_0 \Phi(\bar{\beta}^\pm, \bbz) &= +\left(-\hbar^2\bar\d_+^2 -
 \hbar^2\bar\d_-^2 + 12 (\exp (-4\sqrt{3}\bbp)\right)^{\half}\circ
 \Phi (\bar{\beta}^\pm, \bbz)\cr
 &\equiv \Theta^{\half}\circ\Phi(\bar{\beta}^\pm,\bbz)\ ,\cr}
\ee
where the choice of plus sign in the square-root ensures the
positivity of frequency and where, for notational simplicity, we have
omitted bars on the subscripts of the derivative operators. The
physical inner product is given by:
\be
 \IP{\Psi}{\Phi} = \int_{\bbz= k} \ \rd\bbp\wedge \rd\bbm\
 \overline{\Psi}\Phi \ ,
\ee
where $k$ is a constant, the integral on the right side being
independent of the choice of $k$. Denote the Hilbert space by
$\bar{\cal H}$.  It is obvious that the operator
$-i\hbar\bar\d_0\equiv\sqrt\Theta$ generating the conditional symmetry
is self adjoint on $\bar{\cal H}$, whence the symmetry is unitarily
implemented.

The symmetry thus provides us with a Hilbert space structure on the
space of physical states. To compare this structure with that obtained
in section 4.2, it is convenient to note that the general solution to
(\ref{eq:spiff}) is of the form
\be
 \Phi(\bar\beta^\pm ,\bar\beta^0) = (\exp {i\over\hbar}\Theta^\half
 \bar\beta^0)\circ \phi(\bar\beta^\pm )
\ee
where $\phi(\bar\beta^\pm)$ is in $L^2(R^2, \rd\bar\beta^+ \wedge
\rd\bar\beta^-)$.
There is thus a natural isomorphism between the Hilbert space
$\bar{\cal H}$ of physical states $\Phi(\bar\beta^\pm, \bar\beta^0)$
and the Hilbert space $\bar{\cal H}_0$ of square-integrable (with
respect to measure 1) functions $\phi(\bar\beta^\pm)$.  In section
4.2, we found that the Bianchi II model admits four time-independent
Dirac observables: $\tp_\pm$ and $\tilde{v}_\pm$. Can we express the
corresponding operators on the barred Hilbert space? Using the
definition of the canonical transformation which took us to the tilde
variables, it is straightforward to express the quantum analogs of
three of these observables as operators on the Hilbert space
$\bar{\cal H}_0$. We have:
\bea
 \hat{\tilde{p}}_+\circ \phi(\bar\beta^\pm) &=& (\hat{\bar{p}}{}_+^2
 +12 \exp(-4\sqrt{3}\,\hat{\bar\beta}{}^+))^\half \circ
 \phi (\bar\beta^\pm), \cr
 \hat{\tilde{p}}_-\circ\phi (\bar\beta^\pm) &=& -i\hbar \bar\d_-\phi
  (\bar\beta^\pm),\quad {\rm and},\cr
 \hat{\tilde{v}}_-\circ\phi (\bar\beta^\pm) &=& \bar\beta_-
 \cdot\phi(\bar\beta^\pm) \ .
\eea
We will use the Hilbert space $\bar{\cal H}_0$ and these three operators
thereon in the next sub-section.
The fourth Dirac observable, $\hat{\tilde{v}}_+$, on the other hand, seems
difficult to express as an operator on the barred Hilbert spaces. Its
classical analog is $(\tpp\tbp + (\tpp^2\tbz)/\tpz)$ and, because of
the non-triviality of the canonical transformation from $(\bar\beta^+,
\bar{p}_+)$ to $(\tilde\beta^+ ,\tilde{p}_+)$, its expression in the
barred variables is quite complicated, involving not only square-roots but
also logarithms of polynomials in $\bar\beta^+$ and $\bar{p}_+$. Consequently
one encounters severe factor ordering ambiguities in the passage to quantum
theory. In this sense then we do not have access to a {\it full set} of
time independent Dirac observables on the barred Hilbert spaces. In this
sense, the quantum theory based on conditional symmetries is not as complete
as that constructed in section 4.

\goodbreak
\subsection{Comparison: Time independent Dirac observables}

We now wish to compare the quantum theory obtained in the previous
subsection with that obtained in section 4.2.
In the quantum description just constructed, the Hilbert space $\bar
{\cal H}$ is the space of normalizable positive frequency solutions to
a Klein-Gordon equation with a static potential, while the Hilbert
space $\tilde{\cal H}$ of section 4 is the space of functions on the right
half \halfcone\ of the null cone in momentum space. One's first
impulse may be to conclude that $\bar{\cal H}$ is {\it twice} as large
as $\tilde{\cal H}$ since there is now no trace of the (non-holonomic)
constraint that led us to the {\it half} cone. We will show that this
conclusion is incorrect.

It is convenient for this purpose to recast the mathematical framework
of section 4.2 by emphasizing the role of wave functions
$\phi(\tilde{p}_\pm)$ over those of solutions $\Phi(\tilde{p}_\pm,
\tilde{p}_0)$ to the quantum constraint. (See Eq. (\ref{eq:ps}).) The
Hilbert space $\tilde{\cal H}_0$ of physical states is then $L^2(R^2,
\frac{d\tpp d\tpm}\tpp)$. The four time independent Dirac operators
have the following action on $\tilde{\cal H}_0$:
\bea
 \hat\tpp\circ\psi(\tp_\pm)  &=& \tpp\cdot\psi (\tp_\pm),
 \hphantom{i\d_+} \quad
 \hat\tpm\circ\psi(\tp_\pm)  = \tpm\cdot\psi (\tp_\pm)\cr
 \hat{\tilde{v}}_+\circ \psi(\tp_\pm) &=& i\hbar\tpp \tilde\d_+ \psi(\tp_\pm),
 \quad
 \hat{\tilde{v}}_-\circ \psi(\tp_\pm) = i\hbar \tilde\d_- \psi(\tp_\pm),
\eea
and they provide us with a complete set of quantum observables. To relate
the two quantum theories, we wish to ask if there is a unitary
map from the Hilbert space $\bar{\cal H}_0$ to the Hilbert space
$\tilde{\cal H}_0$ introduced in section 6.1, which interacts in the
correct way with these observables.

To set up such a map, it is easiest to first find in each space a set
of basis vectors corresponding to the same set of commuting Dirac
observables. One commuting set is given by $(\hat\tpp, \hat\tpm)$.
Since the corresponding operators act simply by multiplication on
$\tilde{\cal H}_0$, the spectra are trivial to compute on this Hilbert
space: both operators have a purely continuum spectrum, that of
$\hat\tpp$ is given by $(0, \infty)$ while that of $\hat\tpm$ is given
by $(-\infty,\infty)$. Thus, a simultaneous eigenstate $\ket{\tpp,
\tpm}$ of the two operators is labelled {\it only} by the two real
numbers, $\tpp$ being restricted to be positive; there is no further
degeneracy. The form of the inner product on $\tilde{\cal H}_0$
suggests that we normalize these eigenvectors such that: $\IP{\tpp
,\tpm} {\tpp', \tpm'} = \tpp\delta(\tpp, \tpp') \delta(\tpm,\tpm')$.

What is the situation with $\bar{\cal H}_0$? Since $\tpm \equiv \bar{p}_-$,
the operator $\hat\tpm$  corresponding to $\tpm$ is simply $-i\hbar \bar
\d_-$, whose spectrum is continuous with values in the full range $(-\infty,
\infty)$. The operator $\hat\tpp$ is more complicated.
The canonical transformation defining the tilde variables yields:
\be \label{eq:tp}
 \hat{\tilde{p}}{}_+^2 = \hat{\bar{p}}{}_+^2 + 12 \exp(-4\sqrt{3}\hat{\bar
 \beta}{}^+).
\ee
Since the ``potential'' is positive and goes to zero as $\tbp$ goes to
infinity, the spectrum of the operator $\hat{\tp}{}_+^2$ is continuous
and takes all values between $(0, \infty)$. The key question is
whether the spectrum is degenerate. If so, the spectrum of its
positive square root, $\hat\tpp$, would also be degenerate, and the
two quantum descriptions would be {\it inequivalent}. There is a
standard textbook argument (see, e.g., \cite{ll:qm}) which establishes
the non-degeneracy of the spectra of Hamiltonians in 1-dimensional
potential problems of non-relativistic quantum mechanics. With a small
extension to accommodate the fact that the ``potential'' does not go
to zero (in fact diverges) as $\bar\beta^+$ tends to $-\infty$, this
argument ensures that the spectrum of $\hat{\tp}{}_+^2$, and hence also
that of $\hat\tpp$, is non-degenerate. (This is in sharp contrast to
the spectrum of $\hat{\bar{p}}{}_+^2\equiv -\hbar^2\bar\d_+^2$ which
is obviously 2-fold degenerate; the key difference in the two cases is
the presence of the potential term in (\ref{eq:tp})). Thus, on the
Hilbert space $\bar{\cal H}_0$ as well, the kets $\ket{\tpp ,\tpm}$
provide us with a complete basis, which we choose to be normalized
such that $\IP{\tpp ,\tpm} {\tpp', \tpm'} = \tpp \delta(\tpp, \tpp')
\delta(\tpm,\tpm')$. Since the operator $\hat\tpm$ has the
action $\hat\tpm\phi= -i\hbar
\bar\d_-\phi$, the simultaneous eigenfunctions of $\hat{\tilde{p}}_\pm$
have the functional form: $\IP{\bbp,\bbm}{\tpp,\tpm} =
(1/\sqrt{2\pi}) (\exp \frac{i}\hbar \bar\beta^- \tpm) f_{\tpp} (\bar\beta^+)$,
where $f_{\tpp} (\bar\beta^+)$ are the suitably normalized
eigenfunctions of the operator $\hat\tpp$.

We can now set up the required isomorphism $U : \bar{\cal H}_0\mapsto
\tilde{H}_0$:
\be
 (U\circ\phi)(\tp_\pm) := \frac1{\sqrt{2\pi}}\int d\bar\beta^+ d\bar\beta^-
 (\exp -\frac{i}\hbar\bar\beta^- \tpm) \overline{f_{\tpp}}
 (\bar\beta^+) \phi (\bar\beta^\pm)\ .
\ee
It is straightforward to check that $U$ commutes with the action of the
three time independent Dirac operators, $\hat\tpp, \hat\tpm,
\hat{\tilde{v}}_-$ which are independently defined on the two Hilbert
spaces $\bar{\cal H}_0$ and $\tilde{\cal H}_0$.

So far, the fourth time independent Dirac operator $\hat{\tilde{v}}_+$ is
defined only on $\tilde{\cal H}_0$: We saw in section 6.1 that, if one tries
to define it on $\bar{\cal H}_0$ directly by using the barred operators,
one faces severe factor ordering problems. However, now that we have the
map $U$, we can use it to pull $\hat{\tilde{v}}_+$ back to $\bar{\cal H}_0$
from $\tilde{\cal H}_0$ and simply use the resulting operator $U\circ
\hat{\tilde{v}}_+ \circ U^{-1}$ as the fourth Dirac observable in the barred
quantum theory. (This procedure may be regarded as a solution to the factor
ordering problem.) By construction, then, all four Dirac operators on
$\bar{\cal H}_0$ have the ``correct'' commutation relations among themselves.
When this is done, both the tilde and the barred descriptions are equipped
with a compete set of observables and $U$ provides the isomorphism between
them. In this sense, the two quantum theories are equivalent%
\footnote{
 Note that even though we have fixed the normalization of the basis
 vectors, there is still the freedom to rescale each basis vector by
 a phase factor. The condition that the map should commute with the action
 of the third Dirac observable $\hat{\tilde{v}}_-$ restricts the phase factor
 to depend on $\tpp$ only. Thus, the map $U$ is unique only upto $U\mapsto
 \exp iF(\tilde{p}_+) \cdot U$, for some real-valued function
 $F(\tilde{p}_+)$. If we change the map, the image of $\hat{\tilde{v}}_+$
 on $\bar{\cal H}_0$ will change. However, the equivalence result holds for
 any of these $U$.}.

\goodbreak
\subsection{Deparametrization}

We saw in the previous two subsections that the barred description, by
itself, is not as complete as the tilde description since we have direct
access only to three of the four time independent Dirac observables.
Nonetheless, because the scalar constraint could be recast as a
Schr\"odinger evolution equation (\ref{eq:spiff}), we can still deparametrize
the theory satisfactorily and discuss quantum dynamics. We will first expand
on this observation and then relate the resulting dynamical description
to the one obtained in section 5.

Let us begin with the Hilbert space $\bar{\cal H}$. The operators
$\hat{\bar{\beta}}{}^A$ and $\hat{\bar{p}}_+$ are clearly {\it not}
Dirac operators. Nonetheless, we can follow the procedure of section 5.2
to introduce {\it time dependent} Dirac operators $\hat{\bar{\beta}}{}^A(t)$
and $\hat{\bar{p}}_A(t)$. Again, $\hat{\bar{\beta}}{}^0(t)$ is a multiple of
identity and the true degrees reside in $\hat{\bar{\beta}}{}^\pm$ and
$\hat{\bar{p}}_\pm$. As before, a general time dependent Dirac operator has
the form:
\be
 \left(\hat{F}(\bar\beta^A,\bar{p}_A)(t)\circ\Phi\right)(\bar\beta^A) :=
 e^{\frac{i}\hbar\sqrt{\Theta}({\bar\beta}{}^0 - t)}\circ
 F(\hat{\bar\beta}{}^A,\hat{\bar{p}}_A)\circ \Phi(\bar\beta^\pm,
\beta^0=t)\ .
\ee
Thus, the situation is completely analogous to the one we encountered in
section 4; this discussion of quantum dynamics is quite insensitive to
whether or not one has access to a complete set of time {\it independent}
Dirac operators. However, since the analog of the Hamiltonian $H$ in
(\ref{eq:tddogen}) is now $\sqrt\Theta/\hbar$ and since $\Theta$
involves the complicated potential term (Eq. (\ref{eq:spiff})), the explicit
expressions of the resulting time dependent Dirac operators are now quite
involved and explicit calculations, correspondingly harder. Nonetheless,
{\it in principle}, the conditional symmetry approach provides us with the
machinery needed for the construction of histories of physical interest,
to phrase a variety of dynamical questions and to make physical
predictions. Note also that the inner product makes each of the operators
$\hat{\bar{\beta}}{}^\pm$ and $\hat{\bar{p}}_\pm$ self adjoint; classical
reality conditions {\it are} incorporated properly.

Is there a sense in which this analysis of dynamics on $\bar{\cal H}$ is
equivalent to that on $\tilde{\cal H}$ performed in section 5.3? Thanks
to the map $U$ constructed in the previous subsection, the answer is in the
affirmative. For a general operator, one must make factor ordering choices
on both Hilbert spaces. If the operators are so ordered that, on the
$\bar\beta^0 =0$ and $\tilde\beta^0= 0$ slices, the two operators are
related by the map $U$, i.e., $U\circ\hat{\bar{F}}(t=0)\circ U^{-1} =
\hat{\tilde{F}}(t=0)$, then the two sets of dynamical predictions,
calculated independently on the two Hilbert spaces $\bar{\cal H}$ and
$\tilde{\cal H}$ will coincide. This happens because the map $U$ sends the
Hamiltonian $\hat{\tilde{H}}$ (\ref{eq:derkins}) on $\tilde{\cal H}$ to
the operator $\sqrt\Theta/\hbar$ on $\bar{\cal H}$.  In this sense, the
two deparametrizations are quantum mechanically equivalent. This is
interesting because the deparametrizations are not related to each
other trivially. They do {\it not}\ correspond just to different slicings
of a given domain space of wave functions. Indeed, the domain spaces
are themselves different in the two cases ---one is spanned by the three
$\bar\beta^A$ and the other by the three $\tilde\beta^A$. Because
$\bar\beta^A$ are mixtures of the $\tilde\beta^A$ and their momenta
(and vice versa), there is no simple geometrical relation between the
two domain spaces and hence between the two sets of slicings.

However, it is
important to bear in mind that the difference in the two parametrizations
is of a quite special nature. On the full phase space, $\bar\beta^0=
\tilde\beta^0$, whence the two deparametrizations (i.e., foliations of
the constraint surface) agree classically. A difference arises in the
quantum theory only because the two descriptions result from choosing
different polarizations on the phase space. In a more general situation,
one would encounter a difference in the deparametrization already at the
classical level: there may exist two distinct foliations of the classical
phase space {\it and} two related polarizations such that the quantum
constraint reduces to the Schr\"odinger form with respect to each of them.
It would be extremely valuable to construct and analyse such an example.

\goodbreak
\section{Discussion}

The four main results of this paper are contained in the four sections,
3-6: existence of canonical transformations which removes the potential
term in the dynamical constraint; completion of the quantization program
of \cite{newbook,aa:rst}; extraction of dynamics from a frozen formalism;
and, comparison between two distinct quantization procedures. Their content
and ramifications can be summarized as follows.

First, for a fairly large class of spatially homogeneous models --diagonal,
intrinsically multiply transitive ones-- we exhibited in section 3 a
canonical transformation which removes the ``potential'' term in the scalar
or the Hamiltonian constraint of geometrodynamics. In terms of the new
canonical variables, then, the scalar constraint is purely quadratic in
momenta. Furthermore, the supermetric turns out to be flat! What distinguishes
one model from another is the global topology of the constraint surface, or,
of the effective configuration space. Finally, note that in all but type I
models, the scalar constraint in the ADM variables --and hence the dynamics
of, say, anisotropies-- is quite complicated. It is striking that by using
the new canonical variables, these complications can be bypassed both
classically and quantum mechanically. We effectively map an interacting
problem to a free problem. The solution of the free problem is trivial
and all the physics of the original problem is coded essentially in the
transformation relating the two.

Indeed, we found in section 4 that using the new canonical variables,
one can carry out the general quantization program of \cite{newbook,aa:rst}
to completion in a straightforward fashion. This exercise did, however,
teach us something about the program itself. First, we could
isolate a complete set of Dirac observables and, using the reality
conditions, equip the space of quantum states with a unique Hermitian
structure \cite{rendall:pc}. This result provides an
independent check on the strategy of \cite{newbook,aa:rst} for
obtaining the inner product using reality conditions rather than
attempting to implement symmetries unitarily. In particular, we saw
that it is {\it not} necessary to ``deparametrize'' \cite{kk:qg2} the
theory to obtain the inner product; the issue of constructing a
consistent mathematical framework is thus divorced from the conceptual
difficulties associated with the problem of time
\footnote{This point is important because there exist interesting constrained
 systems --including 2+1 dimensional general relativity on a 2-surface
 with genus $\ge 2$-- where a global deparametrization is either
 difficult or impossible but where the reality conditions suffice to
 select the inner product \cite{newbook,aa:rst}. Note, however, that
 the general strategy does not {\it require} that the Dirac observables
 be time {\it independent}. If one can isolate a time and find a
 complete set of time {\it dependent} Dirac observables (such as
 anisotropies in the DIMT models), one can use them to impose the
 reality conditions with equal ease. This point has been misunderstood
 in some recent reviews of the quantization program.}.
Second, we gained technical insight on the kind of problems that
can arise if the set of observables under considerations fails to be
complete even on a set of measure zero on the classical phase space.

However, many interesting  questions cannot be even phrased purely in
terms of the Dirac observables introduced in section 4.
Indeed, since there is no notion of time, we cannot address the issue
of dynamics. In section 5, we therefore recast the scalar constraint
in another form, that of the Schr\"odinger equation. Thus, one of the
arguments of the physical quantum states is now interpreted as time,
with respect to which other arguments --representing the ``true,
dynamical'' degrees of freedom-- evolve. We could then introduce a
1-parameter family of operators, whose dependence on the parameter
provided us with information about their ``evolution.'' For each value of
the parameter, the operator weakly commutes with the constraints; it
maps the physical states to other physical states and is therefore a
genuine Dirac observable. In this sense, the ``deparametrization''
considered here is ``covariant.'' In particular, a
physical state is simply a solution to the quantum constraint equation
rather than a restriction of the solution to a suitable ``slice'' in
the effective configuration space. We used the newly
introduced, ``time-dependent Dirac observables'' to analyse how
anisotropies evolve and what happens to the classical singularities in
the quantum theory. We found that the singularities persist; minisuperspace
quantization does not remove them. Finally, one can also use
this framework to construct various physically interesting histories
and examine, e.g., whether they decohere. In this sense then, we have all
the machinery needed in the measurement theory of closed systems
\cite{jh:lh}.

The result on persistence of singularities is at first surprising.
After all, dynamics is unitarily implemented by a self-adjoint
Hamiltonian and there is a general belief that unitary evolution can
not lead to any singularities.  We would like to emphasize that this
belief is simply unfounded. To see this point, let us return
momentarily to the classical Hamiltonian description of, say, the
Bianchi I model. In this case, one can find a globally defined,
complete set of constants of motion even though almost every dynamical
trajectory runs into a singularity. If we choose 4 constants of motion
$K_i(\beta^A, p_A)$ and $\beta^0$ as coordinates on the constraint
surface in phase space, each dynamical trajectory is given simply by
$K_i =({\rm const})_i$; there is no trace of a singularity in this
form of the solution. The singularity appears when we examine how
quantities like the anisotropies $\beta^\pm$ or curvature scalars
change along the trajectories. Indeed, even as the trajectory plunges
into the singularity, the values of $K_i$ remain well-defined, equal
to the constants that specify the trajectory. The situation in the
quantum theory is similar. Since $\hat{p}_\pm$ constitute a complete
set of commuting observables which commute with the Hamiltonian,
dynamics is trivialized: as the wave function evolves, the absolute
value of the wave function $\psi(p_\pm, \bz)$ remains constant and
only the phase oscillates as $\exp\beta^0 (i\sqrt{p_+^2 +p_-^2})$. We
see no trace of the singularity: The expectation values of any
observables made out of $\hat{p}_\pm$ remain constants. Indeed, the
expectation values of observables made out of $\hat{p}_\pm$ {\it and}\
$\hat{v}_\pm$ --which together generate the entire algebra of Dirac
observables-- remain {\it finite}. We see the singularity only when we
examine other observables, e.g., the expectation values of
anisotropies or curvature scalars. We see no reason to rule out a
similar circumstance in more general cases; even in the general
context, singularities may persist inspite of unitarity of quantum
evolution. In the DIMT models, however, the situation is more striking
than what may be expected in the general context: in these cases, we
could find {\it complete} sets of observables which are constants of
motion and therefore remain perfectly well behaved through out the
evolution, even when other, (time dependent) observables diverge
---signaling a physical singularity. This is a striking illustration
of the procedure of mapping a non-trivial model to a trivial one. The
non-trivial physics is simply hidden; it does not go away. It can be
uncovered by examining the physically interesting observables of the
original model in the solution of the trivial one.

Finally, we saw in section 6 that in the type II model, a quantum theory
could be constructed following two different avenues: the quantization
program of \cite{newbook,aa:rst}; and the use of conditional symmetries
\cite{kk:qg2,atu:I}. The second approach is closer to traditional quantum
field theories where the Hilbert space structure in the quantum theory is
dictated by the presence and structure of suitable symmetries. We compared
the two approaches in section 6. The framework resulting from the first
approach is more complete in that we have direct access to a complete set of
(time independent) Dirac observables. In the conditional symmetries approach,
it appears very difficult to introduce one of these Dirac observables directly.
However, if we ``pull back'' this last Dirac observable from the Hilbert space
constructed in the first approach, the two descriptions are
equivalent.

Although the models considered here are dynamically non-trivial, the
existence of a multiply transitive isometry group intrinsic to each
spatially homogeneous slice makes them exactly soluble. It is this
solubility that lies at the heart of the canonical transformations.
Therefore, the technical considerations of this paper are not likely to
be useful to the discussion of full quantum gravity in 3+1 dimensions.
Nonetheless, the qualitative lessons learnt here are likely to be
valuable: they provide us with further confidence in the general quantization
program of \cite{newbook,aa:rst}; illustrate the simplifications that can be
caused by judicious canonical transformations; and suggest how one can use
the time dependent Dirac observables to probe dynamics in the setting of a
``covariant deparametrization'', and to analyse physical issues such as the
fate of singularities within the framework of canonical quantization.
\bigskip

\goodbreak
{\bf Acknowledgments}:

We are grateful to Jim Hartle and Jorma Louko for discussions. RST would like
to thank Karel Kucha\v r and Carlo Rovelli for explaining their ideas on
the issue of time and deparametrization.
This work was supported in part by the NSF grants PHY90-16733 and
PHY90-08502, by a grant from the Swedish National Science research
Council and by research funds provided by Syracuse University.

\end{document}